\documentclass[amsmath,amssymb,aps,pra,twocolumn,showpacs,longbibliography]{revtex4-1}
\pdfoutput=1
\PassOptionsToPackage{breaklinks}{hyperref}
\AtBeginDocument{\RequirePackage{hyperref}}

\usepackage{graphicx}
\usepackage{verbatim}

\usepackage[utf8]{inputenc}
\newcommand{\ds}{\displaystyle}
\newcommand{\vp}{\varphi}
\DeclareMathOperator{\tr}{tr}
\newcommand{\vt}{\vartheta}

\DeclareMathOperator{\sinc}{sinc}
\usepackage{amsmath}

\begin{document}
\title{Wigner Functions for the  Pair Angle and Orbital
Angular Momentum}
\author{H.A. Kastrup}
\email{hans.kastrup@desy.de}
\affiliation{DESY, Theory Group, Notkestrasse 85, D-22607 Hamburg, Germany}
\begin{abstract}
  The problem of constructing physically and mathematically
  well-defined Wigner functions for the canonical pair angle $\theta$
  and angular momentum $p$ is solved. While a key element for the
  construction of Wigner functions for the
   planar phase space $\{(q,p) \in \mathbb{R}^2\}$ is the Heisenberg-Weyl
  group, the corresponding group for the
  cylindrical phase space  $\{(\theta,p) \in S^1\times \mathbb{R}\}$ is the
  Euclidean group $E(2)$ of the plane and its unitary
  representations. Here the angle $\theta$ is replaced by
  the pair $(\cos\theta, \sin\theta)$ which corresponds uniquely to the points on
  the unit circle. The main structural properties of the Wigner
  functions for the planar and the cylindrical phase spaces are
  strikingly similar.

 A crucial role plays  the $\sinc$ function which
 provides the interpolation for the discontinuous quantized angular momenta
in terms of the continuous classical ones, in accordance with the famous Whittaker
cardinal function well-known from interpolation and sampling theories.

The quantum mechanical marginal distributions for angle (continuous) and angular
momentum (discontinuous) are as usual uniquely obtained by appropriate integrations
of the $(\theta,p)$ Wigner function. Among the examples discussed is an elementary
system of simple ''cat'' states.
\end{abstract}
\pacs{03.65.Ca, 03.65.Ta, 03.65.Wj, 42.50.Tx}
\maketitle
\section{Introduction}
In 1932 Wigner introduced a function on the classical $(q,p)$ phase
space of a given system with the aim to describe quantum mechanical statistical
properties of that system partially in terms of classical ones
\cite{wig,wig1}. After a slow start the concept has become quite
popular and efficient, e.g.\ in quantum optics
\cite{leo,schl,roe,leo2,aga} and time-frequency analysis \cite{groe,coh}.

In view of the successes of Wigner functions on
the topologically trivial planar phase space $\mathbb{R}^2 $ attempts have been made
to generalize the concept to other phase spaces, in particular to that
of a simple rotator around a fixed axis, its position given by an angle $\theta$
and its angular momentum by a real number $p$, thus having a phase space
corresponding topologically to a cylinder of infinite length,
i.e. $S^1\times \mathbb{R}$\,.

A considerable obstacle for the quantum theory of that space is the
treatment of the angle which has no satisfactory self-adjoint operator
counterpart quantum mechanically \cite{ka}. Nevertheless a
``Hermitian'' angle operator was introduced formally, exponentiated to
a seemingly unitary one, and the classically continuous angular
momentum $p$ was replaced by the discrete quantum mechanical
$l \in \mathbb{Z}$ one, thus quantizing a hybrid ``phase space''
$S^1\times \mathbb{Z}$. This approach was started by Berry \cite{ber}
and Mukunda \cite{muk} and has since been pursued by quite a number of
authors. Typical more recent examples are Refs. \cite{soto,prza,muk2}.

The present paper provides consistant Wigner functions on the classical physical phase space
$S^1\times \mathbb{R}$ including a mathematically satisfactory treatment
of the angle $\theta \in \mathbb{S}^1$.

 We shall see below (Subsec.\ II.B.) where the ``hybrid'' case $S^1\times \mathbb{Z}$
 is to
be placed within the framework proposed here.

Parallel to those attempts to find Wigner functions for systems on the
cylinder its quantum mechanics was described -- in two
different approaches -- surprisingly in terms of unitary representations of the
{\it Euclidean group} $E(2)$ of the plane!  Fronsdal initiated the
first of these two in the framework of the so-called Moyal or $\star$
quantization \cite{fron}, followed by a number of papers by other
authors, e.g.\ \cite{gad,arra}.

In a different approach Isham discussed very convincingly \cite{ish}
the group theoretical quantization of the phase space
$S^1\times \mathbb{R}$ in terms of irreducible unitary representations
of $E(2)$ and its covering groups.  His results could be used in order
to discuss in detail the appropriate quantum mechanics of the
$(\theta,p)$-rotator \cite{ka}.

 The present paper is a continuation of
Ref. \cite{ka}. It provides a mathematically and physically consistent Wigner function
on a cylindrical phase space by utilizing the tools discussed in Ref. \cite{ka}:
Representing the angle $\theta$ of the rotator by the equivalent pair $(\cos \theta,
 \sin \theta)$ introduces - together with the angular momentum $p$ - simultaneously
the 3-dimensional Lie algebra of the Euclidean group $E(2)$ in 2 dimensions. Unitary
 representations
of that group provide the appropriate quantum mechanics of the rotator. Using slightly
modified ideas for constructing ``Wigner operators'' from the Lie algebra of associated
 groups
by Wolf and collaborators \cite{wol1,wol2,wol3} leads to the corresponding operator $V(\theta,p)$ in terms
of the Lie algebra of $E(2)$ (Sec.\ II).   Wigner functions are obtained by calculating matrix elements of
that operator within an irreducible representation of $E(2)$. Matrix elements between two
different
states yield Wigner-Moyal functions \cite{moy} whereas two identical states provide Wigner
functions proper \cite{wig}, the latter being a special case of the former.

The resulting structures and properties (Sec.\ III) have a very remarkable resemblence to those of
the well-known $(q,p)$ case (for reviews see, e.g.\ Refs.\ \cite{wig1,fol,gos}).
An essential feature of the $(\theta,p)$ Wigner-Moyal functions is the central role played
by the sinc function ({\it sinus cardinalis})
\begin{equation} \sinc \pi (p-m) = \frac{\sin \pi (p-m)}{\pi (p-m)},\,\, p \in \mathbb{R},\,
 m \in \mathbb{Z},\label{eq:51} \end{equation}
which interpolates between the discontinuous quantized angular momenta $ m$ ($\hbar = 1)$ by means of the
continuous classical ones $p$ (for a graph of the function \eqref{eq:51} see FIG.1 in Subsection IV.C. below).

In 1915 E.T.\ Whittaker  introduced his - by now famous - {\it
cardinal function} \cite{whi} for the continuous interpolation of functions for which discrete values
are known by using the functions  \eqref{eq:51}. Since then those functions have been playing an increasingly
important role in the fields of interpolation, sampling and signal processing theories (see the reviews
 \cite {mcn,ste1,bu,hig,ste2,ste,vet,eld}).

The time evolution of Wigner-Moyal functions is determined by the Hermitian operator
 $K(\theta,p) = i[H,V(\theta,p)]$ (Sec.\ III.C.).

The quantum mechanical {\it marginal} distributions $|\psi(\theta)|^2$ and $b_m = |c_m|^2$, where the
$c_m$ are the expansion coefficients of the wave function $\psi(\vp)$ with respect to the
basis $e_m(\vp) = \exp( i m \vp),\, m \in \mathbb{Z}$, are obtained - as usual - by integrating the
Wigner function $V_{\psi}(\theta,p)$ over $p$ and $\theta$ respectively. The latter integration
yields a cardinal function $\omega_{\psi}(p)$ from which the probabilities $b_m$ can be extracted
by means of the orthonormality of the $\sinc$ functions \eqref{eq:51} (Sec.\ IV.B.).

Four examples of Wigner functions for certain typical states are discussed in Sec.\ IV.C.:
That of the basis function $e_m(\vp)$, that of the ''cat'' state $(e_1(\vp)+e_{-1}(\vp))/\sqrt{2}$,
 that of ''minimal uncertainty''  states $\psi_e(\vp)$ which lead to the
{\it von Mises} statistical distribution $ |\psi_e(\vp)|^2 = \exp(2s\cos\vp)/I_0(2s)$ and which are
the analogue of the {\it Gaussian} wave packets in the $(q,p)$ case and finally that of thermal
states associated with the Hamiltonian $H = \varepsilon L^2$, where $L$ is the quantum mechanical
counterpart of the classical angular momentum $p$.

\section{The auxiliary role of the Euclidean group $\mathbf{E(2)}$}
The close relationship between the rotator and the group $E(2)$ comes
about as follows \cite{ka}: In order to avoid the problems with a
quantization of the angle $\theta$ itself Louisell and Mackey in 1963
suggested independently \cite{louis,mack} to use the pair
$(\cos\theta,\sin\theta)$ instead, because it is in one-to-one
correspondence to the points on the unit circle and it consists of smooth
bounded $2\pi$-periodic functions the quantization of which appears most
likely  to be easier to handle than that for the angle itself.
\subsection{Classical $\mathbf{E(2)}$ group theory}
The  basic functions
\begin{equation}
  \tilde{h}_1 = \cos\theta,\qquad
  \tilde{h}_2 = \sin\theta,\qquad
  \tilde{h}_3 = p_{\theta} \equiv p
\end{equation}
on the classical phase space
\begin{equation}
  \label{eq:42}
  \mathcal{S}_{\theta,p}
  = \{s=(\theta,p);\,\theta \in \mathbb{R}\,\bmod{2\pi},\, p \in
  \mathbb{R}\,\}
\end{equation}
obey the Poisson brackets
\begin{equation}
  \{\tilde{h}_3,\tilde{h}_1\}_{\theta,p} = \tilde{h}_2,\,
  \{\tilde{h}_3,\tilde{h}_2\}_{\theta,p} =-\tilde{h}_1,\,
  \{\tilde{h}_1,\tilde{h}_2\}_{\theta,p} = 0. \label{eq:3}
\end{equation}
They constitute the Lie algebra of the 3-parametric Euclidean group
$E(2)=\{g(\vt,\vec{a})\}$ of the plane: \\
$\vec{x} \rightarrow R(\vt)\cdot\vec{x}+ \vec{a}$, i.e.
\begin{equation}
  \label{eq:4}
  g(\vt,\vec{a})\circ \vec{x} =
  \begin{pmatrix}
    \cos\vt & -\sin\vt \\
    \sin\vt & \cos\vt
  \end{pmatrix}
  \begin{pmatrix}x_1 \\ x_2\end{pmatrix}
  + \begin{pmatrix} a_1 \\ a_2\end{pmatrix}\,.
\end{equation}
 We write
\begin{equation}
  \label{eq:5}
  \begin{pmatrix}
    \cos\vt & -\sin\vt \\
    \sin\vt & \cos\vt
  \end{pmatrix}
  \begin{pmatrix} x_1 \\ x_2\end{pmatrix}
  \equiv R(\vt)\vec{x} \equiv \vec{x}_{\vt}\,.
\end{equation}
In the following it is advantageous to cast the relations \eqref{eq:4}
into those of $3\times 3$ matrices \cite{vil}, for which group
multiplication etc.\ is implemented by matrix multiplication etc.:
\begin{equation}
  \label{eq:14}
  g(\vt,\vec{a})
  \circ \begin{pmatrix} x_1 \\ x_2\\ 1\end{pmatrix}
  = \left(\begin{array}{ccr}
    \cos\vt & -\sin\vt& a_1 \\
    \sin\vt & \cos\vt & a_2 \\
    0& 0& 1
  \end{array}\right)
  \begin{pmatrix} x_1 \\
    x_2\\ 1\end{pmatrix}\,,
\end{equation}
from which one can read off
immediately the Lie algebra generators
\begin{equation}
   \tilde{L} =
   \left(\begin{array}{ccr}
       0 & -1& 0 \\
       1 &0  &0 \\
       0& 0&0
     \end{array}\right)\!,\,
   \tilde{K}_1 =
   \left(\begin{array}{ccr} 0 &0 & 1 \\
       0 &0  &0 \\
       0& 0&0
     \end{array}\right)\!,\,
   \tilde{K}_2 =
   \left(\begin{array}{ccr}
       0 &0& 0 \\
       0 &0  &1 \\
       0& 0&0
     \end{array}\right)\!.
\end{equation}
They  obey the
commutation relations
\begin{equation}
\label{eq:7}
   [\tilde{L}, \tilde{K}_1] = \tilde{K}_2,~~[\tilde{L}, \tilde{K}_2]
= - \tilde{K}_1,~~[\tilde{K}_1, \tilde{K}_2] = 0~.
\end{equation}
These are obviously isomorphic to those of Eq.\eqref{eq:3}.  In the
corresponding quantum theory the generators $\tilde{L}$ and
$\vec{\tilde{K}} = (\tilde{K}_1,\tilde{K}_2)$ will turn into self-adjoint
 operators $L$
and $\vec{K}$ which describe the quantum mecchanics of the angular
momentum $p$ and the pair $(\cos\theta, \sin \theta)$.

 As to the
latter there is the following subtlety: In Eqs.\eqref{eq:7} the
generators $\tilde{K}_j$ are not normalized: If one multiplies them
with a nonzero constant the new $\tilde{K}_j$ obey the same
communitation relations as the old ones.  This means that the relevant
classical observable here is a \emph{direction,} a \emph{vector} (e.g.\
$\in \mathbb{C}$) or a \emph{half-ray}
\begin{equation}
  \label{eq:31}
  \vec{\chi}\,(\theta) = \chi\,(\cos\theta, \sin\theta)\,,\, \chi > 0\,,
\end{equation}
which carries all the required information on the angle $\theta$!

 In
the corresponding quantum theory the eigenvalues of $\vec{K}$ are
\begin{equation}
  \label{eq:6}
  \vec{k}(\vp) = k\,(\cos\vp,\sin\vp)\,, \, k>0\,,
\end{equation}
with $k^2$ the value of the Casimir operator $\vec{K}^2$; therefore
$k$ is not ``quantized''. It also characterizes the irreducible
unitary representations of $E(2)$ \cite{ish,vil,sug}.  We shall see
below that the modulus $k$ can easily be integrated out with only the
essential pair $(\cos\theta, \sin\theta)$ left.

Using series expansion for the exponentials we get
\begin{equation}
   g(\vt,\vec{a}=0)= g(\vt)= e^{\displaystyle \vt\tilde{L}} =
   \left(\begin{array}{ccr}
       \cos\vt & -\sin\vt& 0 \\
       \sin\vt & \cos\vt & 0 \\
       0& 0&1
     \end{array}\right)\,,
\end{equation}
\begin{equation}
   g(\vt=0,a_1,a_2=0)= g(a_1)=e^{\displaystyle
  a_1\tilde{K}_1}=\begin{pmatrix}1&0&a_1\\0&1&0\\0&0&1 \end{pmatrix}\,,
\end{equation}
\begin{equation}
   g(\vt=0,a_1=0,a_2)= g(a_2)=e^{\displaystyle
  a_2\tilde{K}_2}=\begin{pmatrix}1&0&0\\0&1&a_2\\0&0&1 \end{pmatrix}\,.
\end{equation}
The group element in relation \eqref{eq:14} can then be written as
\begin{equation}
  \label{eq:1}
  g(\vt,\vec{a}) = g(a_2)\circ g(a_1)\circ g(\vt)\,.
\end{equation}
We now  arrive at  a critical part of the paper: Like in the case of the
Heisenberg-Weyl group \cite{fol,gos} the special group element \cite{wol2,chi}
\begin{align}
  \label{eq:2}
  g_0(\vt,\vec{b})& = \exp(b_1\tilde{K}_1 + b_2\tilde{K}_2 +
  \vt\tilde{L})\nonumber\\
  & =
  \begin{pmatrix}
    \multicolumn{2}{c}{R(\vt)}&\sinc(\vt/2)\,R(\vt/2)\vec{b}\\
    0&0&1
  \end{pmatrix}
\end{align}
plays an essential role in the following construction
 of the Wigner
function. Its importance was emphasized previously by Wolf and
collaborators \cite{wol1,wol2,wol3}.

 Comparing Eq.\eqref{eq:2}
and Eq.\eqref{eq:14} shows that the translations are now parametrized
in an angle dependent way. We have
\begin{equation}
  \label{eq:8}
  \vec{a}= \sinc(\vt/2)\,R(\vt/2)\vec{b}\,.
\end{equation}
Crucial for the rest of the paper is that the element $g_0$ from
Eq. \eqref{eq:2} can also be written as
\begin{equation}
  \label{eq:9}
  g_0(\vt,\vec{b}) = e^{\ds (\vt/2)\tilde{L}}\circ e^{\ds
    \sinc(\vt/2)\,\vec{b}\cdot\vec{\tilde{K}}}\circ e^{\ds(\vt/2)\tilde{L}}\,,
\end{equation}
which amounts to a kind of Weyl symmetrization! From Eq. \eqref{eq:8} we then
obtain
\begin{equation}
  \label{eq:10}
 g_0(\vt,\vec{a}) = e^{\ds (\vt/2)\tilde{L}}\circ e^{\ds
    R(-\vt/2)\vec{a}\cdot\vec{\tilde{K}}}\circ e^{\ds(\vt/2)\tilde{L}}\,.
\end{equation}
Now the undesirable factor $\sinc(\vt/2)$ has dropped out.
\newpage
\subsection{Constructing the quantum mechanical \\ Wigner operator}
In the quantum theory the group element \eqref{eq:10} is ``promoted'' to the
unitary operator
\begin{equation}
  \label{eq:11}
  U_0(\vt,\vec{a})=  e^{\ds i (\vt/2)L}\circ e^{\ds i
    \vec{a}_{(-\vt/2)}\cdot\vec{K}}\circ e^{\ds i (\vt/2)L}\,.
\end{equation} (See Eq. \eqref{eq:5} as to notations.)

The unitary operator $U_0$ is supposed to act in an appropriate
Hilbert space in which the Euclidean group is represented unitarily
and in which the operators $L$, $K_1$ and $K_2$ are
self-adjoint. Possible unitary representations are the irreducible,
the quasi-regular and the regular ones \cite{vil}. We here consider
only irreducible unitary representations of $E(2)$
\cite{vil,ish,sug,chi}.

For the construction of appropriate Wigner functions on the phase
space \eqref{eq:42} we have to combine the operator \eqref{eq:11} in a suitable
way with the classical variables $\vec{\chi}(\theta)$ (see
Eq.\eqref{eq:31}) and the angular momentum $p$\,. Following
lessons from the Heisenberg-Weyl group \cite{fol,gos} and from --
appropriately modified -- suggestions by Wolf et al. \cite{wol1,wol2}
the present proposal as to a proper $(\theta,p)$ Wigner operator for
the phase space with the topology $S^1\times \mathbb{R}$ is
\begin{gather}
  \label{eq:13}
   V[\vec{\chi}(\theta),p]=
  \frac{1}{(2\pi)^3}\int_{-\pi}^{\pi}\!d\vt\int_{-\infty}^{\infty}\!da_1da_2\, \hat{U}_0\,, \\
  \hat{U}_0 = e^{\ds i(L-p)(\vt/2)}\circ e^{\ds i(\vec{K}-\vec{\chi}(\theta))\cdot \vec{a}_{(-\vt/2)}}\nonumber \\
 \circ\, e^{\ds i (L-p)(\vt/2)}\,.\nonumber
\end{gather}
This is a kind of ordered group averaging (with invariant Haar measure
$dg(\vt,\vec{a})= d\vt da_1da_2\,$) of the differences between the
classical quantities $\vec{\chi}(\theta)$ and $p$ and their
corresponding operator counterparts.

The ansatz \eqref{eq:13} differs from that of Ref.\ \cite{wol2}
essentially by a different operator ordering which avoids the
obstructive explicit factor $\sinc(\vt/2)$\,.

All irreducible unitary representations of $E(2)$ and its -- infinitely
many -- covering groups can be implemented in a Hilbert space
$L^2(S^1,d\vp/2\pi;\delta)$ with the scalar product
\begin{equation}
  \label{eq:12}
  (\psi_2,\psi_1) =
  \int_{-\pi}^{\pi}\frac{d\vp}{2\pi}\psi_2^{\ast}(\vp)\psi_1(\vp)\,,
\end{equation}
and a basis
\begin{equation}
  \label{eq:15}
  e_{n,\delta}(\vp)=e^{\ds i(n+\delta)\vp}\,,(e_{m,\delta},
  e_{n,\delta})=\delta_{mn},\,m,n \in \mathbb{Z}\,,
\end{equation}
where $\delta \in [0,1)$ characterizes the choice of a covering group
\cite{ka,ish}. A $\delta \neq 0$ becomes important for fractional
orbital angular momenta \cite{ka}. $\delta_{mn}$ is, of course, the
usual Kronecker symbol.

 For any $\psi^{[\delta]}(\vp) \in L^2(S^1,d\vp/2\pi;\delta)$, i.e.
$\psi^{[\delta]}(\vp + 2\pi) = e^{i 2\pi\delta} \psi^{[\delta]}(\vp)$ we have the
expansion
\begin{equation}
  \label{eq:16}
  \psi^{[\delta]}(\vp) = \sum_{n \in \mathbb{Z}}c_{n}\, e_{n,\delta}(\vp)\,,~c_{n } =
  (e_{n,\delta},\psi^{[\delta]})\,.
\end{equation}
The coefficients $c_n$ are independent of $\delta$! That dependence
of $\psi^{[\delta]}(\vp)$ is taken over   by the $e_{n,\delta}(\vp)$.

In the following we put $\delta = 0$. The general case
$\delta \neq 0$ will be briefly discussed in Subsection V.A. below.

The action of the self-adjoint operators $\vec{K}$ and $L$ is given by
($\hbar$ =1)
\begin{equation}
  \label{eq:17}
  \vec{K}\psi(\vp)= k\,(\cos\vp,\sin\vp)\,\psi(\vp)\,,~~L\psi(\vp) =\frac{1}{i}
\partial_{\vp}\psi(\vp)\,,\end{equation}\begin{eqnarray} \label{eq:18}
  e^{\ds i\vec{a}_{(-\vt/2)}\cdot\vec{K}}\psi(\vp)& =&e^{\ds
    i\vec{a}_{(-\vt/2)}\cdot\vec{k}(\vp)}\psi(\vp)\,,\\e^{\ds i\vt L}\psi(\vp)&=&
 \psi(\vp +\vt)\,.\label{eq:77}
\end{eqnarray}
The functions $e_{n,\delta}(\vp)$ obviously are eigenfunctions of $L$
with eigenvalues $n+ \delta$ whereas
$\vec{K}$ acts as a multiplication operator.

Different values of $k$ belong to different irreducible
representations and the Plancherel measure for the Fourier transforms
on $E(2)$ is $k\,dk\,$\cite{vil,sug,chi}.

We here can see what the replacement of the proper classical phase
space $S^1\times \mathbb{R}$ by $S^1 \times \mathbb{Z}$ means (see also
Ref.\ \cite{ka}): it
amounts to putting $k=0$ in Eqs. \eqref{eq:17} and \eqref{eq:18},
i.e. representing the translations of $E(2)$ -- which correspond to the
pair $(\cos\theta,\sin\theta)$ -- by the identity (the translations form a
normal subgroup).  Thus eliminating
the observable ``angle'' altogether! What is left are infinitely many
irreducible representations $\{e^{in\vp}, n \in \mathbb{Z}\}$ of $SO(2)$ alone and the factor
$\mathbb{Z}$ in  $S^1 \times \mathbb{Z}$ corresponds to that dual of $SO(2)$.
The angle $\theta$ then has
to be introduced forcibly ``by hand''!

For the matrix elements
\begin{equation}
  \label{eq:19}
  V^{(k)}_{\psi_2\psi_1}[\vec{\chi}(\theta),p] \equiv
  (\psi_2,V[\vec{\chi}(\theta),p]
\psi_1)
\end{equation}
we obtain from the relations \eqref{eq:18} and \eqref{eq:77}:
\begin{gather}
  \label{eq:20}
  V^{(k)}_{\psi_2\psi_1}[\vec{\chi}(\theta),p] =
  \frac{1}{(2\pi)^4}\int\! dg(\vt,\vec{a})\, d\vp\, e^{\ds - i\omega +
   \vec{k}(\vp) \vec{a}_{(-\vt/2)}}\nonumber\\ \times
 \psi_2^{\ast}(\vp-\vt/2)\psi_1(\vp+\vt/2)\,,\, \omega = \vec{\chi}(\theta)\cdot
\vec{a}_{(-\vt/2)} + p\,\vt \,.
\end{gather}
Introducing polar coordinates $\vec{a} = a\,(\cos\alpha, \sin\alpha)$
yields
\begin{eqnarray}
  \label{eq:21}
   V^{(k)}_{\psi_2\psi_1}&=&\frac{1}{(2\pi)^4}\int
   d\vt\,d\alpha\,ada\,d\vp\,e^{\ds -ip\,\vt}\times\\ && e^{\ds -ia\chi\cos(\theta
     -\alpha+\vt/2)}
 e^{\ds iak\cos(\vp-\alpha+\vt/2)}\times \nonumber \\ &&
\psi_2^{\ast}(\vp-\vt/2)\psi_1(\vp+\vt/2)\nonumber\,.
\end{eqnarray}
Because of the possible expansion \eqref{eq:16} it suffices to take
$\psi_2(\vp) =e_m(\vp)$ and $\psi_1(\vp)=e_n(\vp)$.

 Recalling the
integral representation
\begin{equation}
  J_n(x) = \frac{1}{2\pi}\int_{-\pi}^{\pi}d\beta e^{\ds i(x\,\sin\beta
      -n\, \beta)},\, n \in \mathbb{Z},
\end{equation}
of the Bessel functions \cite{bess}, their (Hankel transform) ''completeness'' relation \cite{mor}
\begin{equation}
  \label{eq:22}
  \int_0^{\infty}da\,a\,J_m(x\,a)\,J_m(y\,a) =\frac{1}{x}\delta(x-y)
\end{equation}
and using the orthonormality \eqref{eq:15} of the $e_m(\vp)$ we
finally get
\begin{eqnarray}
  \label{eq:23}
  V^{(k)}_{m n}(\theta,p)&=&e^{\ds i
    (n-m)\theta}\left(\frac{1}{k}\delta(k-\chi)\right)\times \\ &&\frac{1}{(2\pi)^2}
\int_{-\pi}^{\pi}d\vt\, e^{\ds i [(m+n)/2 -p]\vt}\,. \nonumber
\end{eqnarray}
\section{Main results}
\subsection{The phase space Wigner-Moyal matrix }
Eq.\ \eqref{eq:23} shows explicitly that the scale factor $\chi$ is a redundent
variable and can easily be eliminated: integration (averaging) with
the Plancherel measure $k\,dk$ gives our
\emph{surprisingly simple but powerful main result}
\begin{eqnarray}
 \lefteqn{ V_{m n}(\theta,p) =  \int_0^{\infty}dk\,k
  V^{(k)}_{m n}(\theta,p) =} \nonumber \\
 && =\frac{1}{(2\pi)^2}\,e^{\ds i(n-m)\theta}\int_{-\pi}^{\pi}d\vt\,
  e^{\ds i[(n+m)/2
    -p]\vt} \label{eq:24}\\ &&  =
  \frac{1}{2\pi}\,e^{\ds i(n-m)\theta}\,\sinc\pi[p-(m+n)/2] \,. \label{eq:69}
\end{eqnarray}
The infinite-dimensional Hermitian (Wigner-Moyal) matrix  $V(\theta,p) = (V_{m n}(\theta,p))$ has the properties
\begin{gather}
  \label{eq:25}
\int_{-\infty}^{\infty}dp\,V_{m n}(\theta,p) = \frac{1}{2\pi}\,e^{\ds
  i (n-m)\theta}\,,\\ \int_{-\pi}^{\pi}d\theta\,V_{m n}(\theta,p)=
 \sinc\pi(p-m)\, \delta_{m n}\,,\label{eq:71} \\
 \int_{-\pi}^{\pi}d\theta\int_{-\infty}^{\infty}dp\, V_{m n}(\theta,p) = \delta_{m n}\,,
 \\
 \tr(V_{m n}(\theta,p))= \frac{1}{2\pi} \sum_{n \in \mathbb{Z}}\sinc\pi(p-n) =
\frac{1}{2\pi}, \label{eq:52} \\
 \int_{-\pi}^{\pi}d\theta\int_{-\infty}^{\infty}dp\, V_{k l}(\theta,p)\, V_{m n}(\theta,p)
 =\frac{1}{2\pi}\delta_{kn}\delta_{lm}\,, \label{eq:76}
\end{gather}
\begin{equation}
 | V_{m n}| \le
  \frac{1}{2\pi}\,,
\end{equation}
where the relations
\begin{eqnarray}
  \label{eq:26}
  \sinc\pi x& =& 1 \mbox{ for } x=0\,, \\  |\sinc\pi x|& <& 1 \mbox{ for } x
  \neq 0\,,\nonumber \\ \int_{-\infty}^{\infty}dx\,\sinc\pi(x+a)& =&1, \, a \in \mathbb{R}\,, \\
\frac{1}{2\pi}\sum_{n \in \mathbb{Z}}e^{\ds i n\beta} = \label{eq:70}
  \delta(\beta) & \mbox{ for }& \beta \in [-\pi,\,+\pi],
\end{eqnarray}
have been used. For Eq.\ \eqref{eq:52} see also Ref.\ \cite{pru}.

It follows from Eq.\ \eqref{eq:76} that
\begin{eqnarray}
\int_{-\pi}^{\pi}d\theta\int_{-\infty}^{\infty}dp\,V^{T}(\theta,p)\cdot V(\theta,p)& =& \frac{1}{2\pi}\mathbf{1}\,, \\
\int_{-\pi}^{\pi}d\theta\int_{-\infty}^{\infty}dp\,V(\theta,p)\cdot V^T(\theta,p) &=& \frac{1}{2\pi}\mathbf{1}\,,
\end{eqnarray}
 where $V^{T} (=V^{\ast})$ is the transpose of the matrix $V$ and $\mathbf{1}$ the unit operator in
Hilbert space. Thus, $V$ is Hermitian and even orthogonal, but not unitary!

Another interesting property of $V(\theta,p)$ is
\begin{equation}
\tr[V(\theta_1, p_1)\cdot V(\theta_2, p_2)] = \frac{1}{2\pi}\,\delta(\theta_1 - \theta_2)\,\sinc\pi(p_1 -p_2)\,.
\end{equation}
Integrating this equation over the pair $(\theta_1,p_1)$ (or $(\theta_2,p_2)$)
consistently yields the relation \eqref{eq:52}.

As already mentioned in the Introduction the function  $\sinc \pi(p-m)$ interpolates the discrete
quantum numbers $m$ etc.\ in terms of the continuous classical variable
$p$ (for more details see  the discussion in Section IV below).

 Notice the following difference as to integrations over angles above: whereas
the integration measure in the scalar product \eqref{eq:12}  is normalized as
 $d\phi/(2 \pi)$,
the corresponding measure for the phase space angle variable $\theta$ is $d\theta$.

For expansions
\begin{equation}
  \label{eq:27}
  \psi_j(\vp) = \sum_{n \in \mathbb{Z}}c_n^{(j)}e_n(\vp)\,,\, j=1,2;\,\,\, c_n^{(j)} = (e_n,\psi_j),
\end{equation}
we get the ``Moyal'' function (see Refs. \cite{fol,schl,gos} as to the corresponding $(q,p)$ case)
\begin{gather}
  \label{eq:28}
  V_{\psi_2 \psi_1}(\theta,p) =\sum_{m,n \in
    \mathbb{Z}}c_m^{(2)\ast}V_{m n}(\theta,p)c_n^{(1)}\\ =
  \frac{1}{(2\pi)^2}\int_{-\pi}^{\pi}d\vt\,e^{\ds
    -ip\vt}\psi_2^{\ast}(\theta-\vt/2)\,
  \psi_1(\theta+\vt/2)\,,\nonumber \\
= (\psi_2, V(\theta,p)\psi_1)\,, \nonumber \end{gather} which, according to Eqs.\ \eqref{eq:25} - \eqref{eq:76},
 has the properties
\begin{gather}
\int_{-\infty}^{\infty}dp\,V_{\psi_2\psi_1}(\theta,p) =
\frac{1}{2\pi}\psi_2^{\ast}(\theta)\, \psi_1(\theta)\,, \label{eq:54}\\
\int_{-\pi}^{\pi}d\theta\,V_{\psi_2 \psi_1}(\theta,p) = \sum_{m \in
    \mathbb{Z}}c_m^{(2)\ast}\sinc\pi(p-m)c_m^{(1)}\,, \label{eq:53}\\
\int_{-\infty}^{\infty}dp\int_{-\pi}^{\pi}d\theta
\,V_{\psi_2 \psi_1}(\theta,p) = (\psi_2,\psi_1)\,,\label{eq:29} \\
\int_{-\infty}^{\infty}dp\int_{-\pi}^{\pi}d\theta \, V^{\ast}_{\psi_2 \psi_1}(\theta,p)\,
 V_{\phi_2 \phi_1}(\theta,p) = \label{eq:45} \\ = \frac{1}{2\pi}
(\psi_1,\phi_1)\,(\psi_2,\phi_2)^{\ast}\,.\nonumber
\end{gather}
On the left hand sides of the Eqs. \eqref{eq:54} - \eqref{eq:45} we have integrals
over phase space functions, whereas on the right hand ones we have quantities from
the corresponding Hilbert space, except for Eq.\ \eqref{eq:53} where the function
$\sinc \pi (p-m)$ occurs. However, because of the orthonormality relation
(References in Section IV below)
\begin{equation}
\int_{-\infty}^{\infty}dp \, \sinc\pi(m-p)\,\sinc\pi(n-p) = \delta_{mn}\, \label{eq:48}
\end{equation}
that sinc-function can be eliminated immediately:
\begin{gather}
\int_{-\infty}^{\infty}dp \, \sinc\pi(n-p) \label{eq:55}
 \int_{-\pi}^{\pi}d\theta\,V_{\psi_2 \psi_1}(\theta,p) =  \\
\int_{-\infty}^{\infty}dp \, \sinc\pi(n-p)
\sum_{m \in
    \mathbb{Z}}c_m^{(2)\ast}\sinc\pi(p-m)c_m^{(1)} = \nonumber \\
= c_n^{(2)\ast}\,c_n^{(1)}\,, \nonumber
\end{gather}
a relation which will become important in Sec.\ IV for $\psi_2 = \psi_1$.
\subsection{Phase space functions associated with \\ Hilbert space
 operators}
If $\psi_1(\vp) = A\psi(\vp)$, where $A$ is some operator,
(e.g.\ $\cos\vp,\, \sin\vp,\, L $ or functions of these), then Eq.\ \eqref{eq:29}
 provides  matrix
elements of $A$ and if $\psi_2 =\psi$ its expectation value $(\psi, A \psi)$ {\it in terms of integrals over phase space densities}:

 For $\psi_1 = A\psi$ we
have
\begin{equation}
  \label{eq:34}
  c_n^{(1)} = \sum_{k \in \mathbb{Z}}c_k \,(e_n,Ae_k)
\end{equation}
and it follows from Eq.\ \eqref{eq:29} that
\begin{eqnarray}
  \label{eq:35}
  (\psi_2, A\psi)& =& \int_{-\infty}^{\infty}dp\int_{-\pi}^{\pi}d\theta\, a_{\psi_2\psi}(\theta,p)\,, \\
 a_{\psi_2\psi}(\theta,p)& =& \sum_{m, n \in \mathbb{Z}}\frac{1}{2}c^{(2)\ast}_m [V(\theta,p)
\cdot A+A\cdot V(\theta,p)]_{m n}c_n\,, \nonumber
\end{eqnarray}
which expresses a quantum mechanical matrix element in terms of a
phase space integral over an associated density $a_{\psi_2\psi}(\theta, p)$!
The (Weyl) symmetrization of $V=(V_{m k})$ and $A =(A_{k n})$ in Eq.\ \eqref{eq:35}
is required for $a_{\psi\psi}(\theta, p)$ to be real if $A$ is Hermitian!

For $\psi_2 = e_m, \psi = e_n$  we get
\begin{eqnarray}
  \label{eq:39}
A_{mn}& \equiv &  (e_m,Ae_n) = \int_{-\infty}^{\infty}dp\int_{-\pi}^{\pi}d\theta\,a_{mn}(\theta,p)\,, \nonumber \\
a_{mn}(\theta,p) &=& \frac{1}{2}[V(\theta,p)\cdot A+ A\cdot V(\theta,p)]_{m n}\,,
\end{eqnarray}
which can formally be written as
\begin{equation}
A =(A_{mn}) =  \int_{-\infty}^{\infty}dp\int_{-\pi}^{\pi}d\theta\,\frac{1}{2}[V(\theta,p)\cdot A+ A\cdot V(\theta, p)]\,.\end{equation}
>From the diagonal elements
 \begin{equation}
a_m(\theta,p) = \frac{1}{2}[V(\theta,p)\cdot A+
A\cdot V(\theta,p)]_{m m} \end{equation}
we obtain
\begin{eqnarray}
  \label{eq:36}
  \tr A =&& \sum_{m \in \mathbb{Z}}  \int_{-\infty}^{\infty}dp\int_{-\pi}^{\pi}d\theta\,
 a_m(\theta,p) \\
=&& \int_{-\infty}^{\infty}dp\int_{-\pi}^{\pi}d\theta\, \tr[A\cdot V(\theta,p)]\,.
\nonumber
\end{eqnarray}
If $A$ is a diagonal density matrix
 \begin{equation} \rho =(\lambda_n\,\delta_{mn}),\, \lambda_n \ge 0, \, \sum_{n \in \mathbb{Z}}\lambda_n = 1,
\label{eq:94}
\end{equation} then we have
\begin{eqnarray}\label{eq:84}
&& \tr[\rho\cdot V (\theta,p)] = \frac{1}{2 \pi} \sum_{m \in \mathbb{Z}}
\lambda_m \sinc \pi (p-m)\,, \\
&& \int_{-\infty}^{\infty}dp\int_{-\pi}^{\pi}d\theta\,\tr[\rho\cdot V (\theta,p)]
= \sum_{m \in \mathbb{Z}}\lambda_m = 1\,.
\end{eqnarray}
Using the relations \eqref{eq:48} the probabiities $\lambda_m$ can be extracted from
Eq.\ \eqref{eq:84}:
\begin{equation}
\lambda_m = 2\pi  \int_{-\infty}^{\infty}dp\, \sinc \pi (p-m)\, \tr[\rho\cdot V (\theta,p)]\,.
\end{equation}
As $A$ in Eq.\ \eqref{eq:36} is any (trace class) operator we may take $A= O\cdot\rho $
where $\rho$ is a density
matrix and $O$ a self-adjoint observable. We then have
\begin{equation}
  \label{eq:40}
  \tr (\rho\cdot O) =  \int_{-\infty}^{\infty}dp\int_{-\pi}^{\pi}d\theta\,\frac{1}{2}
\tr [V(\theta,p)\cdot \{O, \rho\} ]\,.
\end{equation}
As $ \tr (\rho\cdot O)= \tr (O\cdot \rho)$ the right-hand side of Eq.\
\eqref{eq:40}
 has to be symmetrized in $\rho$ and $O$, too.

The phase space representation of the trace $\tr(A\cdot B)$ of a product $A\cdot B$
 can also be dealt with in a very similar way as in the $(q,p)$ case
\cite{leo,schl,roe,aga,groo}:

Let us write down $\tr[A\cdot V(\theta,p)]$ explicitly:
\begin{eqnarray}\label{eq:86}
&&\tr[A\cdot V(\theta,p)] = \sum_{m,n \in \mathbb{Z}}A_{mn}V_{nm}(\theta,p)= \\
&&  \frac{1}{(2\pi)^2} \sum_{m,n \in \mathbb{Z}} A_{m n} e^{\ds i(m-n)\theta}
\int_{-\pi}^{\pi}d\vt\,e^{\ds i[(m+n)/2 -p]\vt}\,, \nonumber
\end{eqnarray}
with the corresponding expression for $\tr[B\cdot V(\theta,p)]$.
Inserting these into
\begin{equation}
\int_{-\infty}^{\infty}dp\int_{-\pi}^{\pi}d\theta\,\tr[A\cdot V(\theta,p)]
\tr[B\cdot V(\theta,p)]
\end{equation}
and carrying out the integrations yield the important relation
\begin{equation}
  \tr(A\cdot B) = 2\pi \int_{-\infty}^{\infty}dp\int_{-\pi}^{\pi}d\theta\,
\tr[A\cdot V(\theta,p)]\,
\tr[B\cdot V(\theta,p)]\,.
\end{equation}
Especially we have for the expectation value of the operator $O$
for a given $\rho$:
\begin{eqnarray} \langle O\rangle_{\rho} &=& \tr(\rho\cdot O) \\
 & =& 2\pi \int_{-\infty}^{\infty}dp\int_{-\pi}^{\pi}d\theta\,
\tr[\rho\cdot V(\theta,p)] \nonumber
\tr[O\cdot V(\theta,p)]\,,
\end{eqnarray}
with $\tr[\rho\cdot V(\theta,p)]$ from Eq.\ \eqref{eq:84}.

\subsection{Time evolution}
 Inserting the expansions \eqref{eq:27} into their Schr\"odinger equations
\begin{equation}
  \label{eq:37}
  i\partial_t\psi_j(t,\vp) =
 H\psi_j(t,\vp)
\end{equation}
implies
\begin{equation}
  \label{eq:38}
   i\, \dot{c}^{(j)}_m(t) = \sum_{n \in \mathbb{Z}} c_n^{(j)}(t)(e_m,He_n)\,,
\end{equation}
which leads to the Schr\"odinger time evolution
\begin{equation}
  \label{eq:32}
  \partial_tV_{\psi_2 \psi_1}(\theta,p;t) = i(\psi_2(t),[H,V]
\psi_1(t))\,.
\end{equation}
That is, the time evolution of $V_{\psi_2 \psi_1}$ is determined by the operator (matrix)
\begin{equation}
K(\theta,p) = i[H,V(\theta,p)],\label{eq:89}
\end{equation}
which is Hermitian if $H$ and $V$ are Hermitian and for which $\tr K = 0$.

A simple example is
\begin{equation}\label{eq:93}
H = \varepsilon L^2,~~~~ (e_m, H e_n) = \varepsilon\, n^2 \delta_{mn},
\end{equation}
which implies the matrix
\begin{equation}
(K_{mn}(\theta,p)),~ K_{mn} = (e_m,K e_n) = i\varepsilon (m^2-n^2)V_{mn}(\theta,p).
\end{equation}
The evolution \eqref{eq:32} holds, of course, also for the Wigner function proper
$V_{\psi}(\theta,p;t)$ for which $\psi_2 = \psi_1 = \psi$ in Eq.\ \eqref{eq:28} and which is discussed in the next section.

There it will also be discussed that the Wigner function is given by
\begin{equation}
V_{\rho}(\theta,p;t) = \tr[\rho(t)\cdot V(\theta,p)], \label{eq:88}
\end{equation}
if a state is not characterized by a single wave function $\psi(t)$ but by
a density matrix $\rho(t)$ which obeys the von Neumann equation
(we are still in the Schr\"odinger picture !)
\begin{equation}
\partial_t\rho(t) = -i[H,\rho(t)]\,.
\end{equation}
Inserting this into
\begin{equation}
\partial_t V_{\rho}(\theta,p;t) = \tr[\partial_t\rho(t)\cdot V(\theta,p)]
\end{equation}
yields
\begin{equation}
\partial_t V_{\rho}(\theta,p;t) = \tr[\rho\cdot K(\theta,p)]\,,
\end{equation}
where $K(\theta,p)$ is defined in Eq.\ \eqref{eq:89}.
\newpage

\section{ Wigner function for \\  a given state}
\subsection{General properties}
If $\psi_2 = \psi_1= \psi$ in Eq.\ \eqref{eq:28} then the real function
\begin{eqnarray}
V_{\psi}(\theta,p)&=&  \frac{1}{(2\pi)^2}\int_{-\pi}^{\pi}d\vt\,e^{\ds
    -ip\vt}\psi^{\ast}(\theta-\vt/2)\,
  \psi(\theta+\vt/2)\nonumber\\  \label{eq:44}  &=& \sum_{m,n \in \mathbb{Z}} c_m^{\ast}V_{mn}
(\theta,p)c_n\,  \\ &=& (\psi,V(\theta,p)\psi) \nonumber
\end{eqnarray} is
the strict analogue of the original Wigner function. According to Eqs.
 \eqref{eq:54} - \eqref{eq:45} and \eqref{eq:55}  it obeys
\begin{eqnarray}
\int_{-\infty}^{\infty}dp\,V_{\psi}(\theta,p)& =&
  \frac{1}{2\pi}|\psi(\theta)|^2  \label{eq:90} \\ \int_{-\pi}^{\pi}d\theta \,V_{\psi}(\theta,p)
 & =& \sum_{n \in \mathbb{Z}}|c_n|^2\sinc\pi(p-n)\nonumber \\ & \equiv & \omega_{\psi}(p)\,,\label{eq:62}  \\
 \int_{-\infty}^{\infty}dp\, \omega_{\psi}(p)\,\sinc\pi(p-m) &=& |c_m|^2\,,  \label{eq:47} \\
\int_{-\infty}^{\infty}dp\int_{-\pi}^{\pi}d\theta \,V_{\psi}(\theta,p)& =&\sum_{n
  \in \mathbb{Z}}|c_n|^2 =1\,,\label{eq:46}
\end{eqnarray} \begin{equation}\label{eq:30}
\int_{-\infty}^{\infty}dp\int_{-\pi}^{\pi}d\theta
\,V_{\psi_2}(\theta,p)\,
V_{\psi_1}(\theta,p) = \frac{1}{2\pi}|(\psi_2,\psi_1)|^2\,.
\end{equation}
In addition the inequality
\begin{equation}\label{eq:56}
|V_{\psi}(\theta,p)| \leq 1/\pi\,
\end{equation}
holds. It follows from Schwarz's inequality as follows: the expression \eqref{eq:44} can be written as
\begin{eqnarray} V_{\psi}(\theta,p)& =& \frac{1}{2 \pi} (\chi_2,\chi_1)\,, \\
\chi_1(\theta,p;\vt)& =& e^{-ip\vt/2}\psi(\theta + \vt/2)\,, \nonumber \\
\chi_2(\theta,p;\vt)& =& e^{ip\vt/2}\psi(\theta - \vt/2)\,. \nonumber
\end{eqnarray}
Because
\begin{equation}
|(\chi_2,\chi_1)|^2 \leq (\chi_2,\chi_2)\,(\chi_1,\chi_1)
\end{equation}
and
\begin{eqnarray}
(\chi_1,\chi_1)& =& \int_{-\pi}^{\pi}\frac{d\vt}{2\pi} |\psi(\vt/2 + \theta)|^2 \\
&=& 2\int_{-\pi/2}^{\pi/2}\frac{d\beta}{2\pi} |\psi(\beta)|^2 \nonumber \\
&\leq& 2 (\psi,\psi) = 2, \nonumber
\end{eqnarray}
with the same for $(\chi_2, \chi_2)$,
the inequality \eqref{eq:56} follows.

Eq.\ \eqref{eq:44} gives the Wigner function for a pure state $\psi$. It can immediately
be generalized to a mixed state characterized by a density matrix $\rho$:

 First we rewrite
the scalar product $\eqref{eq:44}$ as the trace of two operators: If $P_{\psi}$ is the projection
operator onto the state $\psi$ $(P_{\psi}= |\psi\rangle \langle \psi|$ in Dirac's notation),
 then its matrix elements are given by
\begin{equation}
(P_{\psi})_{mn}=c_m\,c_n^{\ast}\,.
\end{equation}
Therefore Eq.\ \eqref{eq:44} can be written as
\begin{equation}
V_{\psi}(\theta,p) =(\psi,V(\theta,p)\psi) = \tr[P_{\psi}\cdot V(\theta,p)]\,.\label{eq:110}
\end{equation}
As $P_{\psi}$ is a special density matrix the generalization of $V_{\psi}$ to a mixed
state with density matrix $\rho$ is obviously
\begin{equation}\label{eq:99}
V_{\rho}(\theta,p) = \tr[\rho\cdot V(\theta,p)]\,,
\end{equation}
a quantity we know already from the last Section.

Again using the relations \eqref{eq:25} - \eqref{eq:76} we now have, instead of Eqs.\
\eqref{eq:90} - \eqref{eq:30}:

\begin{align}
\int_{-\infty}^{\infty}dp\,V_{\rho}(\theta,p)& =
  \frac{1}{2\pi} \sum_{m,n \in \mathbb{Z}} e_m(\theta)\rho_{mn}e_n^{\ast}(\theta)\,, \label{eq:108} \\
 \int_{-\pi}^{\pi}d\theta \,V_{\rho}(\theta,p)
 & = \sum_{m \in \mathbb{Z}} \rho_{mm}\sinc\pi(p-m)\nonumber\\ &\equiv  \omega_{\rho}(p)\,,  \end{align} \begin{gather}
 \int_{-\infty}^{\infty}dp\, \omega_{\rho}(p)\,\sinc\pi(p-m) = \rho_{mm}\,,\label{eq:107}  \\
\int_{-\infty}^{\infty}dp\int_{-\pi}^{\pi}d\theta \,V_{\rho}(\theta,p) =\sum_{m
  \in \mathbb{Z}}\rho_{mm} =\tr(\rho) =1\,, \\
\int_{-\infty}^{\infty}dp\int_{-\pi}^{\pi}d\theta
\,V_{\rho_2}(\theta,p)\,
V_{\rho_1}(\theta,p) = \frac{1}{2\pi}\tr(\rho_2\cdot \rho_1)\,.
\end{gather}
Eq.\ \eqref{eq:107} gives only the diagonal elements of the density matrix $\rho$. Its generalization follows immediately
from the relation \eqref{eq:76}:
\begin{equation}
\int_{-\infty}^{\infty}dp\int_{-\pi}^{\pi}d\theta\,V_{kl}(\theta,p)\,V_{\rho}(\theta,p) = \frac{1}{2\pi}\,\rho_{kl},\label{eq:109}
\end{equation}
of which Eq.\ \eqref{eq:107} is a special case.

The representations of the Wigner functions \eqref{eq:110} and \eqref{eq:99} as traces make their invariance under
unitary transformations explicit, especially under those of Eqs.\ \eqref{eq:18} and \eqref{eq:77}.

\subsection{Marginal densities}
For the Wigner function $W_{\psi}(q,p)$ on the {\em classical} phase space  $\{ (q,p) \in  \mathbb{R}^2\}$
 one has the  marginal {\em quantum mechanical}  distributions
 \begin{eqnarray} \int dp\,W_{\psi}(q,p) &=&  |\psi(q)|^2, \label{eq:58} \\
 \int dq\,W_{\psi}(q,p) &=& |\hat{\psi}(p)|^2 \label{eq:59}
\end{eqnarray}  for $q$ and $p$ separately. The properties \eqref{eq:58} and \eqref{eq:59} constitute one
of the main requirements the Wigner function $W_{\psi}(q,p)$ should fulfil \cite{wig1,leo,schl,roe,aga,fol,gos}.

At first sight those properties do hold in our case only  for the
marginal density $|\psi(\theta)|^2$ in Eq.\ \eqref{eq:90}.  However, the situation here is only slightly
more complicated, but in principle as straightforward as in the $(q,p)$ case:
The quantum mechanical marginal probabilities for the quantized angular momentum are  the discontinuous
numbers
 \begin{equation} b_m = |c_m|^2\,,~m \in \mathbb{Z},  \label{eq:41}    \end{equation}
whereas the $\theta$-integration in Eq.\ \eqref{eq:62} yields ''only'' the function $\omega_{\psi}(p)$.

 But  Eq.\
\eqref{eq:47} shows how the marginal probabilities \eqref{eq:41} can be extracted uniquely from $\omega_{\psi}(p)$.

The density $\omega_{\psi}(p)$ of Eq.\ \eqref{eq:62} is an example of  Whittaker's famous {\em cardinal
function} $C(f,h)(x)$ associated with  a function $f(x)$ the values of which are known only  for a discrete subset of the otherwise continuous arguments $x$ \cite{whi},
 an important case well-known
  from interpolation, sampling and signal processing theories
 (see the reviews \cite{mcn,ste1,bu,hig,ste2,ste,vet,eld}):
\begin{equation}
C(f,h)(x) = \sum_{n \in \mathbb{Z}} f(n\,h)\,\sinc[\pi(x-nh)/h]\,,
\end{equation}
where $h$ is the step size.

In our case we have
\begin{eqnarray}
  C(g,h=1)(p)& =&  \sum_{m \in \mathbb{Z}}[g(p=m)= b_m]\sinc\pi(m-p)
 \nonumber \\ &=& \omega_{\psi}(p) \label{eq:49}
\end{eqnarray} for a possible function $g(p), p \in \mathbb{R},$ which now interpolates  the different discrete values $|c_m|^2$,
i.e.\ $g(p=m) = |c_m|^2$.
 The interval $h$ between  adjacent  supporting points here has the value $h=1$.
If the Fourier transform $\hat{f}(u)$ of $f(x)$  is ``band-limited'', i.e. $\hat{f}(u)$ vanishes outside the interval $u \in [-\pi/h, \pi/h]$ (Palais-Wiener case)
then one even has $C(f,h)(x) = f(x)$ (Whittaker-Shannon sampling theorem, mathematically proved by Hardy \cite{har}).

The  functions $\sinc\pi(p-m), m \in \mathbb{Z}$, of Eq.\ \eqref{eq:48} form a complete orthonormal basis for the Hilbert space of such functions $g(p)$ on
the real line $\mathbb{R}$ (see also Ref.\ \cite{chr}).

Here then is the complete set of marginal densities from above (Eqs.\ \eqref{eq:90}, \eqref{eq:47}, \eqref{eq:108} and \eqref{eq:109}),
 for a wave function $\psi(\vp) = \sum_{n \in \mathbb{Z}} c_n\,e_n(\vp)$ or a density matrix
 $\rho = (\rho_{mn})$:
\begin{eqnarray}
&&\frac{1}{2\pi} |\psi(\theta)|^2 = \int_{-\infty}^{\infty}dp\,V_{\psi}(\theta,p), \\
&& |c_n|^2 =  \int_{-\infty}^{\infty}dp \, \int_{-\pi}^{\pi}d\theta\,\sinc\pi(p-n) V_{\psi}(\theta,p), \\
&& \frac{1}{2\pi} \sum_{m,n \in \mathbb{Z}} \rho_{mn}e_m(\theta)e_n^{\ast}(\theta) =
 \int_{-\infty}^{\infty}dp\,V_{\rho}(\theta,p), \\
&&\frac{1}{2\pi} \rho_{mn} = \int_{-\infty}^{\infty}dp\int_{-\pi}^{\pi}d\theta\,V_{mn}(\theta,p)\,V_{\rho}(\theta,p).
\end{eqnarray}
\subsection{Examples}
\subsubsection{Wigner function of the basis function $e_m(\vp)$}
\begin{figure}[htb]
\includegraphics{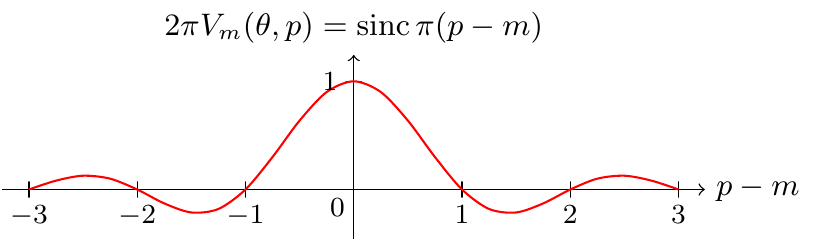}
\caption{\label{fig:emvp}The Wigner function $V_{m}(\theta,p)$ - Eq.\ \eqref{eq:50} - for the basis function
$e_m(\vp)$}
\end{figure}

If $c_m=1$ and $c_n=0$ for $n\neq m$ then
the diagonal matrix element
 \begin{equation} V_{m}(\theta,p) =(1/2\pi) \sinc\pi(p-m) \label{eq:50}\end{equation}
is the Wigner function of the  basis function $e_m(\vp)$. The corresponding
graph is shown in FIG.~\ref{fig:emvp}.

The Wigner function \eqref{eq:50} is independent of $\theta$ and can become negative as a function
of $p$, e.g.\ in the interval $\pm (p-m) \in (1,2)$. So a Wigner function on $S^1 \times \mathbb{R}$
can have negative values, too, like in the $(q,p)$ case:
 The Wigner function (phase space density) $W_n(q,p)$
for the $n$-th energy level of the quantized harmonic oscillator has the value $\lambda (-1)^n,\,\lambda > 0,$
at $(q=0,p=0)$ \cite{schl1}.
In the $ (q,p)$ case negativity of  Wigner functions on certain subsets of the classical phase space is  interpreted
as a consequence of quantum effects. The same interpretation applies to the Wigner function here:
Whereas the classical angular momentum $p$ is continuous its quantized counterpart is discrete. This is reflected
by the properties of the Wigner function \eqref{eq:50}. The difference between the $(q,p)$ case and the $(\theta,p)$
one is that  in the former case the basic variables $q$ and $p$ in general remain continuous in the quantum theory, too,
whereas in the latter the basic angular momentum variable $p$ becomes discontinuous.

The integrals \eqref{eq:90} - \eqref{eq:46} become trivial here:
\begin{eqnarray}
\int_{-\infty}^{\infty}dp\,V_{m}(\theta,p)& =&
  \frac{1}{2\pi}  \\ \label{eq:33} \int_{-\pi}^{\pi}d\theta \,V_{m}(\theta,p)
 & =& \sinc\pi(p-m)\\ & =& \omega_{m}(p)\,, \nonumber \\
\int_{-\infty}^{\infty}dp\, \omega_{m}(p)\,\sinc(p-m)& =&|c_m|^2 = 1  \,,  \\
\int_{-\infty}^{\infty}dp\int_{-\pi}^{\pi}d\theta \,V_{m}(\theta,p)& =&1.
\end{eqnarray}
\subsubsection{Wigner functions of simple ``cat'' states}
The eigenstates $e_{+1}(\vp)$ and $e_{-1}(\vp)$ of the angular momentum operator $L$
with eigenvalues $m=\pm 1$ are the first excited states of a rotator with the simple
Hamiltonian $H = \varepsilon \,L^2$, both with the same energy eigenvalue $\varepsilon$. They
represent, however, two different types  of rotations: clockwise and counterclockwise.
The Wigner functions of their simple superpositions (``cat'' states)
\begin{eqnarray} f_{+1,-1:+}(\vp) &=& \frac{1}{\sqrt{2}}[e_{m=1}(\vp)+ e_{m= -1}(\vp)] \\
&=& \frac{1}{\sqrt{2}}[e^{i\vp} + e^{-i\vp}] = \sqrt{2}\cos \vp, \nonumber \\
| f_{+1,-1:+}(\vp)|^2& =& 1+ \cos 2\vp = 2\cos^2\vp,  \label{eq:60} \\
 f_{+1,-1:-}(\vp) &=& \frac{1}{\sqrt{2}}[e_{m=1}(\vp)- e_{m= -1}(\vp)] \label{eq:61} \\
&=& \frac{1}{\sqrt{2}}[e^{i\vp} - e^{-i\vp}] = i \sqrt{2}\sin \vp, \nonumber \\
| f_{+1,-1:-}(\vp)|^2& =& 1- \cos 2\vp = 2 \sin^2\vp, \end{eqnarray}
show in an interesting manner the influence of the interference or entanglement term $\cos 2 \vp$
for the corresponding phase space densities $V_{f_{\pm}} (\theta,p)$ (in the following we write
$f_+(\vp)$ for $ f_{+1,-1:+}(\vp)$).

Inserting $f_+(\vp)$ into Eq.\ \eqref{eq:44} we get its Wigner function
\begin{equation}
2\pi V_{f_+}(\theta,p) = \cos 2\theta \sinc\pi p + \frac{1}{2}[\sinc\pi (p+1) +
\sinc\pi (p-1)]. \label{eq:57}
\end{equation}
The integrals \eqref{eq:90} - \eqref{eq:46} now take the form
\begin{eqnarray}
\int_{-\infty}^{\infty}dp\,V_{f_+}(\theta,p)& =&
  \frac{1}{2\pi}(\cos2\theta +1)\,,  \\  \int_{-\pi}^{\pi}d\theta \,V_{f_+}(\theta,p) \nonumber
 & =& \frac{1}{2}[ \sinc\pi(p+1) + \sinc\pi(p-1)]\\ & \equiv & \omega_{f_+}(p)\,,  \end{eqnarray}
\begin{eqnarray}
\int_{-\infty}^{\infty}dp\, \omega_{f_+}(p)\,\sinc\pi(p+1)& =&|c_{-1}|^2 = \frac{1}{2}  \,,  \\
\int_{-\infty}^{\infty}dp\, \omega_{f_+}(p)\,\sinc\pi(p-1)& =&|c_{+1}|^2 = \frac{1}{2}  \,,  \\
\int_{-\infty}^{\infty}dp\int_{-\pi}^{\pi}d\theta \,V_{f_+}(\theta,p)& =&1.
\end{eqnarray}

\begin{figure}[htb]
\includegraphics{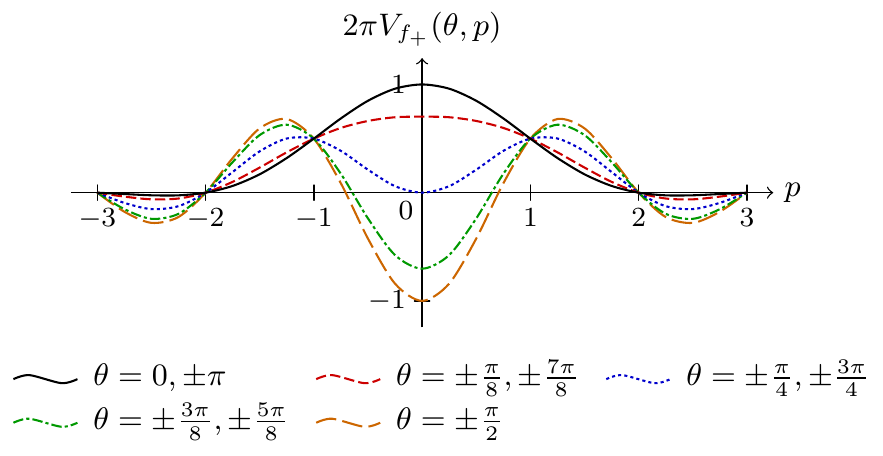}
\caption{\label{fig:cat} Graphs of the Wigner function $2\pi V_{f_+}(\theta,p) = \cos 2\theta \sinc\pi p + \frac{1}{2}[\sinc\pi (p+1) +
\sinc\pi (p-1)]$
  for the ``cat'' state
  $ f_{+}(\vp) = \frac{1}{\sqrt{2}}[e_{m=1}(\vp)+ e_{m=
    -1}(\vp)]$
  as a continuous function of the classical angular momentum $p$ for
  different discrete values of the phase space angle
  $\theta \in [-\pi, +\pi]$. Notice the special cases
  $2\pi V_{f_+}(\theta,p=0) = \cos 2\theta$ and
  $2\pi V_{f_+}(\theta,p = \pm 1) = 1/2 $ .}
\end{figure}

The $\theta$-dependent term in Eq.\ \eqref{eq:57}  represents the {\em interference or entanglement}
part of the probability density \eqref{eq:60}.
Graphs of the function \eqref{eq:57} -- shown in FIG.~\ref{fig:cat} --, parametrized by different angles $\theta \in [-\pi,+\pi]$
 demonstrate the strong influence that interference term has on the phase space function $V_{f_+}(\theta,p)$.

The Wigner function $V_{f_-}$ for the state \eqref{eq:61} is
\begin{eqnarray}
2\pi V_{f_-}(\theta,p)& =& - \cos 2\theta \sinc\pi p \\ && + \frac{1}{2}[\sinc\pi (p+1) +
\sinc\pi (p-1)]. \nonumber
\end{eqnarray}
and can be discussed in the same way as that of $f_{+}$.

Both, $f_+(\vp)$ and $f_-(\vp)$ are special cases of $f_{\alpha}(\vp) = 1/\sqrt{2}(e^{i\vp} + e^{-i\alpha}\,e^{-i\vp})$,
which leads to the same Wigner function as in Eq.\ \eqref{eq:57}, with $2 \theta$ replaced by $2 \theta + \alpha$.
\subsubsection{Minimal uncertainty states}
The following example of a special state is taken from Sec.\ III of Ref.\ \cite{ka} (for a related later discussion see Ref.\ \cite{soto1}): If $A$ and $B $
are two (non-commuting) self-adjoint operators and if the state $\psi$ belongs to their common domain of definition,
then the general ``uncertainty relation'' \cite{rob,schr,jack}
\begin{equation}
(\Delta A)^2_{\psi}(\Delta B)^2_{\psi} \geq |\langle S_{\psi}(A,B)\rangle_{\psi}|^2 + \frac{1}{4}|\langle[A,B]\rangle_{\psi}|^2\,,
\label{eq:64} \end{equation}
holds, where
\begin{eqnarray}
(\Delta A)^2_{\psi} &=& \langle (A-\langle A\rangle_{\psi})^2\rangle_{\psi}\,,\langle A \rangle_{\psi} = (\psi,A \psi),~~~~ \\
S_{\psi}(A,B) &=& \frac{1}{2}(AB+BA) - \langle A \rangle_{\psi} \langle B \rangle_{\psi}.
\end{eqnarray}

Of special interest are those states $\psi =\psi_e$ for which the inequality \eqref{eq:64} becomes an equality.
That equality holds iff $\psi_e$ obeys the linear dependence equation
\begin{equation}\label{eq:65}
(B-\langle B \rangle_{\psi_e})\psi_e = \sigma\, (A- \langle A \rangle_{\psi_e})\psi_e\,,~ \sigma = \gamma + i s\,,
\end{equation}
where the given real parameters $\gamma$ and $s$ determine the following statistical quantities:
\begin{eqnarray}
 \frac{\langle S_{\psi_e}(A,B)\rangle_{\psi_e}}{(\Delta A)^2_{\psi_e}} &=& \gamma \,, \\
 \frac{1}{2i}\frac{\langle [A,B] \rangle_{\psi_e}}{(\Delta A)^2_{\psi_e}} &=& s\,, \\
 \frac{(\Delta B)^2_{\psi_e}}{(\Delta A)^2_{\psi_e}} &=& |\sigma| = \sqrt{\gamma^2+ s^2} \,.
\end{eqnarray}
In the usual case $A=Q, B = P$ the solutions $\psi_e$ of Eq.\ \eqref{eq:65} are  Gaussian wave packets (coherent states),
for which $\gamma = 0$:
\begin{eqnarray}
\psi_e(x)& =& (s/\pi)^{1/4}e^{\ds -(s/2)(x-x_e)^2} e^{\ds i p_e x},\, s > 0,~~~\label{eq:92} \\
&& x_e = \langle Q \rangle_{\psi_e},\, p_e = \langle P \rangle_{\psi_e}\,. \nonumber
\end{eqnarray}
In our case we take $A = S = \sin\vp,\, \langle S\rangle_{\psi_e}= 0$ and $B = L = (1/i)\partial_{\vp}$ (in Ref.\ \cite{ka} Sec.\ III starts with $A= C = \cos\vp$,
but $A=\sin\vp$ is closer to the standard {\em von Mises statistical distribution} \cite{for,mar}).
Assuming here $\gamma = 0$, too, the  choice  $A = \sin\vp$ and  $B = L $ yields the following normalized solution of Eq.\ \eqref{eq:65}:
\begin{eqnarray}
\psi_e(\vp) &=& \frac{1}{\sqrt{I_0(2s)}}e^{\ds i p_e \vp   + s \cos\vp}\,, \, s > 0\,,\label{eq:66}\\
|\psi_e(\vp)|^2& =& \frac{e^{\ds 2s\cos\vp}}{I_0(2s)}\,, \label{eq:68} \\
p_e &=& \langle L \rangle_{\psi_e}\,, \\
\langle S \rangle_{\psi_e} &=& 0\,, \\
\langle C \rangle_{\psi_e} &=& \frac{I_1(2s)}{I_0(2s)}\,.
\end{eqnarray}
If one would allow for $S_e = \langle S \rangle_{\psi_e} \neq 0$ in Eq.\ \eqref{eq:65} the solution \eqref{eq:66}
would include a non-periodic factor $\exp(s\,S_e\vp)$ which cannot be incorporated into any Hilbert space with a scalar product
\eqref{eq:12} and a basis \eqref{eq:15}. Similarly the solution \eqref{eq:92} of Eq.\ \eqref{eq:65} is not square-integrable
on the real line for $ s <0$.

 The functions $I_n(x)$ are modified Bessel functions (for more details which we do not need here see Ref.\
\cite{ka}). The following integral representation for $n \in \mathbb{Z}$  has been und will be used \cite{wat}:
\begin{equation}
I_n(z) = \frac{1}{2\pi} \int_{-\pi}^{\pi} d\phi\, e^{\ds z \cos\phi + i n \phi} = I_{-n}(z)\,.\label{eq:74}
\end{equation}
$I_n(z)$ is real for real $z$ and $I_n(-z) = (-1)^n I_n(z)$.

For the shapes of the distribution \eqref{eq:68} for different values of $s$ see Fig.\ 3.1 of Ref.\ \cite{mar}.

As \begin{equation}
\psi_e(\vp + 2 \pi) = e^{\ds i2\pi p_e} \psi_e (\vp)\,,
\end{equation}
the function \eqref{eq:66} is generally not periodic, because $p_e$ can be any real number.
 However its treatment can be reduced to the case discussed
in the context of Eqs. \eqref{eq:15} and \eqref{eq:16}: we decompose the real number $p_e$ uniquely into
an integer $n_e$ and a fractional rest:
\begin{equation}
p_e = n_e + \delta\,,~ \delta \in [0,1)\,,
\end{equation}
so that
\begin{equation}
\psi_e(\vp) = \frac{1}{\sqrt{I_0(2s)}} e^{\ds i (n_e + \delta)\vp   + s \cos\vp}\,.
\end{equation}
This yields the expansion coefficients
\begin{eqnarray}\label{eq:67}
\lefteqn{c_m = (e_{m,\delta}, \psi_e)} \\& =& \frac{1}{\sqrt{I_0(2s)}}\int_{-\pi}^{\pi}\frac{d\vp}{2\pi}e^{\ds -i(m-n_e)\vp + s\cos\vp} \nonumber \\
& =&  \frac{1}{\sqrt{I_0(2s)}} I_{m-n_e}(s)\,.\nonumber
\end{eqnarray}

In order to calculate the Wigner function of the state \eqref{eq:66} we have to use the Wigner-Moyal matrix
\eqref{eq:63} of the next Section: We then get
\begin{equation}V^{[\delta]}_{\psi_e}(\theta,p) = \sum_{m.n \in \mathbb{Z}} c^{\ast}_m  V^{[\delta]}_{m n}(\theta,p) c_n \,.
\end{equation}
Using for $c_m$, $ c_n$ and $ V^{[\delta]}_{m n}(\theta,p)$ the integral representations \eqref{eq:67} and \eqref{eq:63}
below and observing the relation \eqref{eq:70} (twice) yields the integral representation
\begin{eqnarray}
\lefteqn{~~~ V^{[\delta]}_{\psi_e}(\theta,p)
 =} \label{eq:72}  \\ && \nonumber = \frac{1}{(2\pi)^2 I_0(2s)}\int_{-\pi}^{\pi}d\vt \, e^{\ds - i(p -p_e)\vt + 2s \cos\theta \cos(\vt/2)}
 \\ && \nonumber = \frac{1}{2\pi^2 I_0(2s)}\int_{0}^{\pi}d\vt \, \cos[(p-p_e)\vt] e^{\ds  2s \cos\theta \cos(\vt/2)}, \\
&& ~~~  p_e = n_e + \delta\,. \nonumber
\end{eqnarray}
We drop the label $[\delta]$ of $V^{[\delta]}_{\psi_e}(\theta,p)$ in the following.

\begin{figure}
\includegraphics{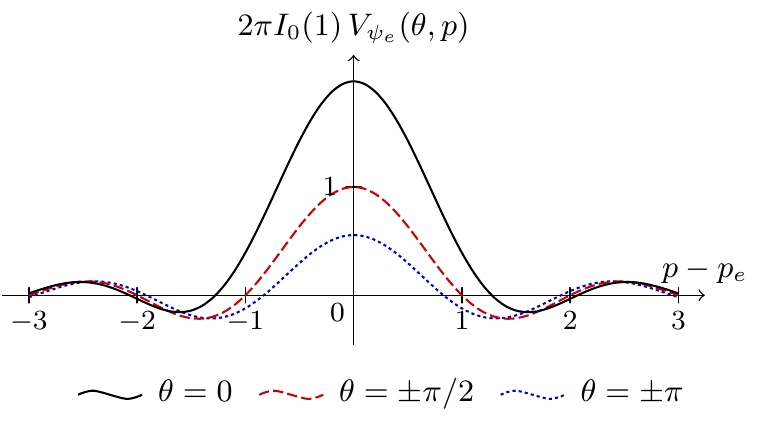}
\caption{\label{fig:vmi} Graphs of the Wigner function \eqref{eq:72} for the
state \eqref{eq:66} with $s = 1/2$. The curves show $2\pi I_0(1) V_{\psi_e}(\theta,p)$ for $\theta = 0,\,\pm \pi/2,\,\pm \pi$ as a function of $p-p_e$.
Numerically: $I_0(1) = 1.2661$ \cite{wa2}.}
\end{figure}
For the values $\theta = \pm \pi/2$ we have
\begin{equation} V_{\psi_e}(\theta = \pm \pi/2,p) =\frac{1}{2\pi I_0(2s)}\,\sinc\pi(p-p_e)\,. \label{eq:91}
\end{equation}
The variables $p$ and $p_e$ are to be treated independently ($p_e$ is the given``observable'' $ \langle L \rangle_{\psi_e}$, $p$ a phase space variable).
The  curves in FIG.  \ref{fig:vmi} have their main maximum for $p = p_e$.

The  Wigner function \eqref{eq:72} indeed obeys the general relations \eqref{eq:90} - \eqref{eq:46} explicitly:

Integrating Eq.\ \eqref{eq:72} over $p$ gives a delta function $\delta (\vt)$ which makes the $\vt$-integral trivial. The result is
the expected marginal density \eqref{eq:68}:
\begin{equation}
\int_{-\infty}^{\infty} dp\,V_{\psi_e}(\theta,p) = \frac{1}{2\pi I_0(2s)} e^{\ds 2 s \cos\theta} = \frac{1}{2\pi} |\psi_e(\theta)|^2.
\end{equation}
The $\theta$-integration and the ensuing determination of $|c_m|^2$ are slightly more subtle:
\begin{eqnarray}
\lefteqn{\int_{-\pi}^{\pi}d\theta\,V_{\psi_e}(\theta,p) = \omega_{\psi_e}(p) =} \\ && = \frac{1}{2\pi I_0(2s)}\,\int_{-\pi}^{\pi}d\vt \,e^{\ds -i(p-p_e)\vt}
I_0(2 s \cos(\vt/2)  \nonumber \\ && =  \frac{1}{\pi I_0(2s)}\,\int_{0}^{\pi}d\vt \,\cos[(p-p_e)\vt]
I_0(2 s \cos(\vt/2), \nonumber
\end{eqnarray}
where again the relation \eqref{eq:74} has been used.

In order to extract from $\omega_{\psi_e}$ the marginal probabilities $|c_m|^2$ - according to Eq.\ \eqref{eq:47} -
we have to multiply $\omega_{\psi_e}(p)$ with $\sinc\pi(p - m - \delta)$ and integrate over $p$:
\begin{equation}
\int_{-\infty}^{\infty} dp\, \omega_{\psi_e}(p)\sinc\pi(p - m - \delta)\,.\label{eq:75}
\end{equation}
Inserting
\begin{equation}
 \sinc\pi(p-m-\delta) = \frac{1}{2\pi} \int_{-\pi}^{\pi}d\alpha\,e^{\ds i(p-m-\delta)\alpha}
\end{equation}
into Eq.\ \eqref{eq:75} yields a delta function $\delta(\alpha - \vt)$ and
\begin{eqnarray}
&&\frac{1}{2\pi I_0(2s)} \int_{-\pi}^{\pi}d\vt\, e^{\ds i(m - n_e)\vt} I_0(2s
 \cos(\vt/2)\\ & =&  \frac{1}{\pi I_0(2s)} \int_{0}^{\pi}d\vt\,
 \cos[(m - n_e)\vt]\, I_0(2s \cos(\vt/2) \nonumber
\end{eqnarray}
for the integral \eqref{eq:75}. As \cite{erd,gra}
\begin{equation}
 \int_{-\pi/2}^{\pi/2}d\beta\,
 \cos(2n \beta)\, I_0(2a \cos\beta) = \pi I_n^2(a)
\end{equation}
the integral \eqref{eq:75} gives indeed the same probability
\begin{equation}
|c_m|^2 = \frac{1}{I_0(2s)}\,|I_{m-n_e}(s)|^2
\end{equation}
as obtained from Eq.\ \eqref{eq:67}.
\subsubsection{Thermal states}
Let  simple rotators with the Hamiltonian \eqref{eq:93} be in a heat bath of temperature $T$.
Then the density matrix \eqref{eq:94} has the form
\begin{eqnarray}\label{eq:96}
\rho& =& (\lambda_n \delta_{mn})\,,~\lambda_n = \frac{e^{\ds -n^2 \varepsilon \beta}}{Z(\beta)},\\
\beta& =& 1/(k_B T), \nonumber \\
Z(\beta) &=& \sum_{n \in \mathbb{Z}} e^{\ds -n^2 \varepsilon \beta}.\label{eq:95}
\end{eqnarray}
The partition function \eqref{eq:95} can be expressed in terms of a $\vt$-function \cite{thet}:
\begin{eqnarray}\label{eq:97}
\vt_3(z, q= e^{\ds i \pi\tau})& \equiv& \vt_3(z|\tau) = \sum_{n \in \mathbb{Z}} q^{\ds n^2} e^{\ds 2ni z} \\
& = & 1+ 2 \sum_{n=1}^{\infty}q^{\ds n^2} \cos 2nz \label{eq:98}\\
&=&  \vt_3(-z,q)\,, \nonumber \\
&& \Im(\tau) > 0\,. \nonumber
\end{eqnarray}
The function \eqref{eq:97} is an {\it entire function} of $z$, real valued for real and imaginary $z$
if $q$ is real.  For real $q$ there are also no zeros on the real and imaginary $z$-axis
and $\vt_3(z,q)$ is positive there.

We now can write
\begin{eqnarray}
Z(\beta)& =& \vt_3(z=0, q = e^{\ds -\varepsilon \beta})\\
& =& \vt_3(z=0|\tau = i \varepsilon \beta/\pi). \nonumber
\end{eqnarray}
For $\varepsilon \beta \gg 1$ (low temperatures) the first order in $q$ in the series \eqref{eq:98}
gives a reasonable approximation:
\begin{equation}\label{eq:102}
Z(\beta) \approx 1 + 2 e^{\ds -\varepsilon \beta},~~1/Z(\beta) \approx 1 - 2 e^{\ds-\varepsilon \beta},~ \varepsilon \beta \gg 1.
\end{equation}
(For $\varepsilon \approx O(\mbox{1 eV})$ and $T \approx 300^{\circ}$K one has $\varepsilon\,\beta \approx 40$).
A corresponding high-temperature approximation can be obtained with the help of Jacobi's famous identity
\begin{equation}\label{eq:103}
\vt_3(z|\tau) = (-i\tau)^{\ds -1/2}e^{\ds -iz^2/(\pi\tau)}\vt_3(z/\tau|-1/\tau).
\end{equation}
It yields
\begin{equation}\label{eq:105}
Z(\beta) = \left(\frac{\pi}{\varepsilon \beta}\right)^{1/2}\vt_3(z=0,q=e^{\ds-\pi^2/(\varepsilon\beta)}),
\end{equation} which for $  \varepsilon \beta \ll 1$ provides the approximation
\begin{eqnarray}
Z(\beta)& \approx& \left(\frac{\pi}{\varepsilon \beta}\right)^{1/2}(1+2e^{\ds-\pi^2/(\varepsilon \beta)}),~\varepsilon \beta \ll 1\,,~~~~~ \\
1/Z(\beta)& \approx& \left(\frac{\varepsilon \beta}{\pi}\right)^{1/2}(1-2e^{\ds -\pi^2/(\varepsilon \beta)}).
\end{eqnarray}
Due to $\varepsilon\beta \ll 1$ and the exponent $\pi^2$ in $q = \exp(-\pi^2/(\varepsilon\beta))$
this $q$ is negligible compared to $1$.
For the Wigner function \eqref{eq:99} we get - according to Eq.\ \eqref{eq:84} - :
\begin{eqnarray}\label{eq:100}
V_{\rho}(\theta,p) &=& \tr(\rho\cdot V(\theta,p)) \\
&=& \frac{1}{2\pi\,Z(\beta)}\sum_{n \in \mathbb{Z}} e^{\ds -n^2 \varepsilon\beta}\sinc\pi(p-n) \nonumber \\
&=&  \frac{1}{(2\pi)^2\,Z(\beta)}\int_{-\pi}^{\pi}d\alpha\, e^{\ds-ip\alpha} \sum_{n \in \mathbb{Z}} e^{\ds -n^2 \varepsilon\beta}
e^{\ds in\alpha}. \nonumber \end{eqnarray} This result obviously can be expressed in terms of the $\vt$-function \eqref{eq:97}, too:
\begin{eqnarray}\label{eq:101}
 V_{\rho}(\theta,p) &=& \tr(\rho\cdot V(\theta,p)) \\ &=& \frac{1}{(2\pi)^2\,Z(\beta)}\int_{-\pi}^{\pi}d\alpha\, e^{\ds -ip\alpha} \vt_3( \alpha/2, q =e^{\ds -\varepsilon \beta})
\nonumber \\
&=&  \frac{1}{2\pi^2\,Z(\beta)}\int_{0}^{\pi}d\alpha\, \cos p\alpha\, \vt_3( \alpha/2, q =e^{\ds -\varepsilon \beta}). \nonumber
\end{eqnarray}
Integrating Eq.\ \eqref{eq:100}, or \eqref{eq:101} respectively, over $\theta$ gives a factor $2\pi$, integrating
 in addition over $p$ gives 1. Multiplying  equation \eqref{eq:100} by $\sinc\pi(p-m)$,
integrating over $p$ and using the relation \eqref{eq:48} gives the  density matrix elements $\lambda_n$ of Eq.\ \eqref{eq:96},
multplied by $1/(2\pi)$.

In order to get some more insights into the properties of the Wigner function \eqref{eq:101} it helps to look at its low- and
high-temperature limits mentioned above:

In the low-temperature limit $\varepsilon \beta \gg 1 $ we have for $\vt_3(\alpha/2, q = \exp(-\varepsilon \beta))$ in first
order of $q$:
\begin{equation}
\vt_3(\alpha/2,q= e^{\ds-\varepsilon \beta}) \approx 1 +2e^{\ds-\varepsilon\beta}\cos\alpha.
\end{equation}
Inserting this into the integral \eqref{eq:101}, observing that \cite{pru2}
\begin{equation}
\int_0^{\pi}d\alpha\,\cos\alpha\,\cos p\alpha =  -\frac{\pi}{2}\,\sinc\pi p \left( \frac{p}{p+1} + \frac{p}{p-1}\right)
\end{equation}
 and using the approximation \eqref{eq:102} yields
\begin{equation}\label{eq:106}
V^{(l)}_{\rho}(\theta,p) = \frac{1}{2\pi} \sinc\pi p\left[1-e^{\ds-\varepsilon \beta}\left(2+\frac{p}{p+1} + \frac{p}{p-1}\right)\right].
\end{equation}
This low-temperature approximation of $V_{\rho}(\theta,p)$ is dominated by the $\sinc\pi p$ function in the neighbourhood of $p =0$. That function also compensates
the poles at $p = \pm 1$: For, e.g.\ $p= 1+ \epsilon$ we have $\lim_{\epsilon \to 0}\sinc\pi(1+\epsilon)/\epsilon = -1$.

 Thus, at very low temperatures the wave function \eqref{eq:106} dominantly describes thermally very
small angular momenta, as expected!

For the high-temperature approximation we use the relation
\begin{eqnarray}\label{eq:104}
&& \vt_3(\alpha/2,q=e^{\ds-\varepsilon\beta})\\ && =\left(\frac{\pi}{\varepsilon\beta}\right)^{1/2}e^{\ds-\alpha^2/(4\varepsilon\beta)}\vt_3\left(z=\frac{i\pi\alpha}{2\varepsilon\beta},
q=e^{\ds -\pi^2/(\varepsilon\beta)}\right), \nonumber
\end{eqnarray} which follows from Jacobi's identity \eqref{eq:103}.

  If the functions $Z(\beta)$ and $\vt_3$ of  Eqs.\ \eqref{eq:105} and \eqref{eq:104} are inserted in Eq.\
\eqref{eq:101} their common prefactor drops out. We then get for $V_{\rho}$ the high-temperature approximation
\begin{equation}
V^{(h)}_{\rho}(\theta,p) = \frac{1}{2\pi^2}\int_0^{\pi}d\alpha\,\cos p\alpha\, e^{\ds-\alpha^2/(4\varepsilon\beta)}.
\end{equation}
As $\varepsilon\beta \ll 1$ the Gaussian under the integral is short-ranged and therefore we can extend the
upper limit of the integral from $+\pi$ to $+\infty$ and obtain \cite{gra2}
\begin{equation}
V^{(h)}_{\rho}(\theta,p) \approx \frac{ \sqrt{\pi\varepsilon\beta}}{2\pi^2}e^{\ds-\varepsilon\beta\,p^2}.
\end{equation}
Thus, at high temperatures the Wigner function \eqref{eq:101} becomes a Boltzmann distribution for the classical angular momenta $p$!
(Note that $\varepsilon p^2$ is the classical counterpart to the quantum mechanical Hamiltonian \eqref{eq:93}.)
Integrating this $V^{(h)}_{\rho}$ over $p$ gives
\begin{equation}
\int_{-\infty}^{\infty}dp V^{(h)}_{\rho}(\theta,p) = \frac{1}{2\pi}.
\end{equation}
The denominator $2\pi$ on the right-hand side is cancelled by the final (trivial) integration over $\theta$.
\section{The  cases  $\mathbf{\delta \neq 0}$ and $\mathbf{\hbar \neq 1}$ }
\subsection{$\mathbf{\delta \neq 0}$}
In Ref.\ \cite{ka} a number of physical examples were mentioned for which the parameter $\delta$
of Eqs.\ \eqref{eq:12} - \eqref{eq:16} is nonvanishing. Another example with $\delta \neq 0$
was discussed in the last Subsection.
 It is, therefore, of interest to indicate the
main changes of the principle formulae in Section III if $\delta \neq 0$:

Going through the  arguments of Subsection II.B, above, now using the basis functions \eqref{eq:15},
we get instead of Eqs.\ \eqref{eq:24} and \eqref{eq:69}
\begin{eqnarray}
 \lefteqn{ V^{[\delta]}_{m n}(\theta,p) } \label{eq:63}  \\
 && =\frac{1}{(2\pi)^2}\,e^{\ds i(n-m)\theta}\int_{-\pi}^{\pi}d\vt\,
  e^{\ds i[(n+m +2\delta)/2
    -p]\vt} \nonumber \\ && =
  \frac{1}{2\pi}\,e^{\ds i(n-m)\theta}\,\sinc\pi[p-(m+n + 2\delta)/2] \label{eq:73} \,.
\end{eqnarray}
The relations \eqref{eq:25} - \eqref{eq:76} now take the form
\begin{gather}
  \label{eq:78}
\int_{-\infty}^{\infty}dp\,V^{[\delta]}_{m n}(\theta,p) = \frac{1}{2\pi}\,e^{\ds
  i (n-m)\theta}\,,\\ \int_{-\pi}^{\pi}d\theta\,V^{[\delta]}_{m n}(\theta,p)=
 \sinc\pi(p-m-\delta)\, \delta_{m n}\,,\label{eq:79} \\
 \int_{-\pi}^{\pi}d\theta\int_{-\infty}^{\infty}dp\, V^{[\delta]}_{m n}(\theta,p) = \delta_{m n},\,\\
 \,\,\tr(V^{[\delta]}_{m n})= \frac{1}{2\pi} \sum_{n \in \mathbb{Z}}\sinc\pi(p-n-\delta) =
\frac{1}{2\pi}, \label{eq:80} \\
 \int_{-\pi}^{\pi}d\theta\int_{-\infty}^{\infty}dp\, V^{[\delta]\ast}_{k l}(\theta,p)\, V^{[\delta]}_{m n}(\theta,p)
 =\frac{1}{2\pi}\delta_{k m}\delta_{l n}\,. \label{eq:81}
\end{gather}
Using expansions \eqref{eq:16} we get instead of Eq.\ \eqref{eq:28}
\begin{gather}
 \label{eq:82}
  V^{[\delta]}_{\psi_2 \psi_1}(\theta,p) =\sum_{m,n \in
    \mathbb{Z}}c_m^{(2)\ast}V^{[\delta]}_{m n}(\theta,p)c_n^{(1)}\\ =
  \frac{1}{(2\pi)^2}\int_{-\pi}^{\pi}d\vt\,e^{\ds
    -ip\vt}\psi_2^{[\delta]\ast}(\theta-\vt/2)\,
  \psi^{[\delta]}_1(\theta+\vt/2)\,\nonumber \\
= (\psi^{[\delta]}_2, V(\theta,p)\psi^{[\delta]}_1)\, \nonumber \end{gather}
The relation \eqref{eq:48} is to be replaced by
\begin{equation}
\int_{-\infty}^{\infty}dp \, \sinc\pi(m+\delta-p)\,\sinc\pi(n+ \delta-p) = \delta_{mn}\,.
\end{equation}
It is apparently rather obvious how one has to proceed if one passes from a
Hilbert space with $\delta = 0$ to one with $\delta \neq 0$, e.g. in Section IV,
Subsections A and B.
\subsection {$\hbar \neq 1$}
In the Sections above we have put $\hbar =1$. To make $\hbar$ explicit again in all the formulae
is easier here than in the $(q,p)$ case, because the basic variable $\theta$ is dimensionless
and only the canonically conjugate angular momentum $p$ which has the dimension
[action] has to be rescaled:

 In order to reintroduce $\hbar$ into the
above formulae the following two replacements are necessary: First the angular momentum operator
$L$ of Eq.\ \eqref{eq:17} has to be rescaled:
\begin{equation}
L = \frac{1}{i}\partial_{\vp} \to \hat{L} = \frac{\hbar}{i}\partial_{\vp}\,.
\end{equation}
Notice that the basis functions \eqref{eq:15} remain unchanged and are now eigenfunctions of
$\hat{L}$ with eigenvalues $\hbar\, n$.

 In addition the classical phase space variable $p$, which
we have treated as dimensionless above, has to be replaced by $p/\hbar$ if $p$ is now interpreted
as a variable with the dimension [action].

 As an example consider the $\sinc$ function of
Eq. \eqref{eq:51}:
\begin{eqnarray}
\sinc \pi(p-m) &\to& \sinc[ \pi (p - \hbar\, m)/\hbar] \label{eq:83}\\
 &=& \hbar\,\frac{\sin[\pi(p- m\,\hbar)/\hbar]}{\pi(p-\hbar\,m)}. \nonumber
\end{eqnarray}
Thus, the $\sinc$-function \eqref{eq:83} is dimensionless.
So are the matrix $V(\theta,p)$ and the wave functions \eqref{eq:27}.
As
\begin{equation}
\lim_{\epsilon \to 0} \frac{\sin(\pi x/\epsilon)}{\pi x} = \delta(x)\,,
\end{equation}
\newline
the associated classical limit of the rescaled expression \eqref{eq:83} can be obtained as
\begin{equation}
\label{eq:43}
 \lim_{\hbar \to 0}\frac{1}{\hbar}\sinc[\pi(p - \hbar\, m)/\hbar] = \delta(p-\hbar\,m)\,,\,\, \hbar\, m = \text{const.}\,.
\end{equation}
In Subsection III.C. the time derivatives $\partial_t$ have to be replaced by $\hbar\, \partial_t$.

\section{remarks}
\subsection{Dirac notation}
Throughout the whole text above I have avoided the use of the widespread Dirac
notion of ''bra'', ''ket''  etc. for describing quantum mechanical states and
operators. The reason being (see, e.g.\ Ref.\ \cite{ka}) that {\it there is no mathematically well-defined angle
operator $\hat{\vp}$ with eigenfunctions $|\vp\rangle$ such that}
\begin{equation}
\hat{\vp}|\vp \rangle = \vp\,|\vp\rangle\,.\label{eq:85}
\end{equation}
The use of such mathematically non-existent objects might be helpful heuristically
if one is appropriately careful: Here the formal objects $|\vp\rangle$ only make
sense in the combination
\begin{equation}
\langle \vp|n\rangle =e^{\ds in\vp} = e_n(\vp)\,,~~L|n\rangle = n |n\rangle\,,
\end{equation}
where $L$ is the well-defined angular momentum operator.
Armed with this mental reservation one may write for the operator $V$ of Eq.\
\eqref{eq:24}
\begin{equation}
\tilde{V}(\theta,p) =\frac{1}{(2\pi)^2}\int_{-\pi}^{\pi}d\vt\,e^{\ds -ip\vt}|
\theta-\vt/2
\rangle \langle \theta + \vt/2|\,,
\end{equation}
the matrix elements $\tilde{V}_{mn}=\langle m|\tilde{V}|n\rangle$ of which are  the same as those in Eq.\ \eqref{eq:24}.

The Wigner function in Eq.\ \eqref{eq:44} may be written as
\begin{equation}
V_{\psi}(\theta,p) =     \tilde{V}_{\psi}(\theta,p) = \langle \psi|\tilde{V}(\theta,p)|\psi \rangle\,,
\end{equation}
where $|\psi\rangle$ is to be considered as an expansion in terms of the
eigenstates $|n\rangle$.

Another interesting example is (see Section III.B.)
\begin{eqnarray}
\tr[V(\theta,p)\cdot A]& =& \tr[\tilde{V}(\theta,p)\cdot A] \label{eq:87} \\
&=&\frac{1}{(2\pi)^2}\int_{-\pi}^{\pi}d\vt\,e^{\ds -ip\vt}\langle \theta + \vt/2|A|
\theta -\vt/2 \rangle\,. \nonumber
\end{eqnarray}
Inserting the completeness relations  $\sum_{m \in \mathbb{Z}}|m\rangle \langle m|=
\mathbf{1}$ and
 $\sum_{n \in \mathbb{Z}}|n\rangle \langle n|=
\mathbf{1}$
before and after the operator $A$ in the last expression leads back to the
results of Section III.B.
\subsection{A possible generalisation and some  related work}
The approach discussed above for phase spaces of the topological type $S^1 \times \mathbb{R}$ can
be generalized, e.g. to a free rigid body in 3 dimensions with one point fixed. Its configuration space
 can be identified with the group $SO(3)$ \cite{chi1,mars}. Its 2-fold covering $SU(2)$ has the topology of $S^3$.
 The associated phase space $S^3 \times \mathbb{R}^3$ can be quantized  in terms of unitary representations
of the Euclidean group $E(4)$ \cite{ish1}.

At a late stage of the present investigations  I became aware of the work by Leaf \cite{leaf1,leaf2} in which the operator
$\Delta(q,p)$ plays a role with respect to the $(q,p)$ Wigner function which corresponds to that of the operator/matrix
$V(\theta,p)$ in the discussions above. Leaf's approach was used by de Groot and Suttorp in their textbook \cite{groo}.

A Wigner-Moyal function on the circle very similar in structure to the one in Eq.\ \eqref{eq:28} above can be found
in Refs.\ \cite{wol4,muk2}. There, however, the classical angular momentum is treated as discontinuous, contrary to  its properties
in its classical phase space and its treatment in the present paper!

\begin{acknowledgments} I am again very grateful to the DESY Theory
  Group for its continuous generous hospitality since my retirement
  from the Institute for Theoretical Physics of the RWTH Aachen.

I thank Krzysztof Kowalski for drawing my attention first to Ref.\ \cite{wol4} after a shorter Letter version
of the present paper had appeared (arXiv:1601.02520v2) and  to Ref.\ \cite{muk2} after the completion of the present paper.
I thank Hartmann R\"omer for a critical reading of the manuscript
and the suggestion to include thermal states among the examples in Sec.\ IV.
\end{acknowledgments}
Finally I am indebted to my son David for providing the figures and for his help with the final \LaTeX\ typewriting.
\bibliography{wignerfunction}

\begin{thebibliography}{65}%
\makeatletter
\providecommand \@ifxundefined [1]{%
 \@ifx{#1\undefined}
}%
\providecommand \@ifnum [1]{%
 \ifnum #1\expandafter \@firstoftwo
 \else \expandafter \@secondoftwo
 \fi
}%
\providecommand \@ifx [1]{%
 \ifx #1\expandafter \@firstoftwo
 \else \expandafter \@secondoftwo
 \fi
}%
\providecommand \natexlab [1]{#1}%
\providecommand \enquote  [1]{``#1''}%
\providecommand \bibnamefont  [1]{#1}%
\providecommand \bibfnamefont [1]{#1}%
\providecommand \citenamefont [1]{#1}%
\providecommand \href@noop [0]{\@secondoftwo}%
\providecommand \href [0]{\begingroup \@sanitize@url \@href}%
\providecommand \@href[1]{\@@startlink{#1}\@@href}%
\providecommand \@@href[1]{\endgroup#1\@@endlink}%
\providecommand \@sanitize@url [0]{\catcode `\\12\catcode `\$12\catcode
  `\&12\catcode `\#12\catcode `\^12\catcode `\_12\catcode `\%12\relax}%
\providecommand \@@startlink[1]{}%
\providecommand \@@endlink[0]{}%
\providecommand \url  [0]{\begingroup\@sanitize@url \@url }%
\providecommand \@url [1]{\endgroup\@href {#1}{\urlprefix }}%
\providecommand \urlprefix  [0]{URL }%
\providecommand \Eprint [0]{\href }%
\providecommand \doibase [0]{http://dx.doi.org/}%
\providecommand \selectlanguage [0]{\@gobble}%
\providecommand \bibinfo  [0]{\@secondoftwo}%
\providecommand \bibfield  [0]{\@secondoftwo}%
\providecommand \translation [1]{[#1]}%
\providecommand \BibitemOpen [0]{}%
\providecommand \bibitemStop [0]{}%
\providecommand \bibitemNoStop [0]{.\EOS\space}%
\providecommand \EOS [0]{\spacefactor3000\relax}%
\providecommand \BibitemShut  [1]{\csname bibitem#1\endcsname}%
\let\auto@bib@innerbib\@empty
%</preamble>
\bibitem [{\citenamefont {{Wigner}}(1932)}]{wig}%
  \BibitemOpen
  \bibfield  {author} {\bibinfo {author} {\bibfnamefont {E.}~\bibnamefont
  {{Wigner}}},\ }\bibfield  {title} {\enquote {\bibinfo {title} {On the quantum
  correction for thermodynamic equilibrium},}\ }\href@noop {} {\bibfield
  {journal} {\bibinfo  {journal} {Phys. Rev.}\ }\textbf {\bibinfo {volume}
  {40}},\ \bibinfo {pages} {749} (\bibinfo {year} {1932})}\BibitemShut
  {NoStop}%
\bibitem [{\citenamefont {Hillery}\ \emph {et~al.}(1984)\citenamefont
  {Hillery}, \citenamefont {O'Connell}, \citenamefont {Scully},\ and\
  \citenamefont {Wig\-ner}}]{wig1}%
  \BibitemOpen
  \bibfield  {author} {\bibinfo {author} {\bibfnamefont {M.}~\bibnamefont
  {Hillery}}, \bibinfo {author} {\bibfnamefont {R.~F.}\ \bibnamefont
  {O'Connell}}, \bibinfo {author} {\bibfnamefont {M.~O.}\ \bibnamefont
  {Scully}}, \ and\ \bibinfo {author} {\bibfnamefont {E.~P.}\ \bibnamefont
  {Wig\-ner}},\ }\bibfield  {title} {\enquote {\bibinfo {title} {Distribution
  functions in physics: Fundamentals},}\ }\href@noop {} {\bibfield  {journal}
  {\bibinfo  {journal} {Phys. Rep.}\ }\textbf {\bibinfo {volume} {106}},\
  \bibinfo {pages} {121} (\bibinfo {year} {1984})}\BibitemShut {NoStop}%
\bibitem [{\citenamefont {Leonhardt}(1997)}]{leo}%
  \BibitemOpen
  \bibfield  {author} {\bibinfo {author} {\bibfnamefont {U.}~\bibnamefont
  {Leonhardt}},\ }\href@noop {} {\emph {\bibinfo {title} {Measuring the Quantum
  State of Light}}},\ \bibinfo {series} {Cambridge Studies in Modern Optics},
  Vol.~\bibinfo {volume} {22}\ (\bibinfo  {publisher} {Cambridge University
  Press},\ \bibinfo {address} {Cambridge, UK},\ \bibinfo {year}
  {1997})\BibitemShut {NoStop}%
\bibitem [{\citenamefont {{Schleich}}(2001)}]{schl}%
  \BibitemOpen
  \bibfield  {author} {\bibinfo {author} {\bibfnamefont {W.~P.}\ \bibnamefont
  {{Schleich}}},\ }\href@noop {} {\emph {\bibinfo {title} {{Quantum Optics in
  Phase Space}}}}\ (\bibinfo  {publisher} {Wiley-VCH},\ \bibinfo {address}
  {Berlin},\ \bibinfo {year} {2001})\BibitemShut {NoStop}%
\bibitem [{\citenamefont {{R{\"o}mer}}(2009)}]{roe}%
  \BibitemOpen
  \bibfield  {author} {\bibinfo {author} {\bibfnamefont {H.}~\bibnamefont
  {{R{\"o}mer}}},\ }\href@noop {} {\emph {\bibinfo {title} {{Theoretical
  Optics: An Introduction}}}},\ \bibinfo {edition} {2nd}\ ed.\ (\bibinfo
  {publisher} {Wiley-VCH},\ \bibinfo {address} {Weinheim},\ \bibinfo {year}
  {2009})\BibitemShut {NoStop}%
\bibitem [{\citenamefont {Leonhardt}(2010)}]{leo2}%
  \BibitemOpen
  \bibfield  {author} {\bibinfo {author} {\bibfnamefont {U.}~\bibnamefont
  {Leonhardt}},\ }\href@noop {} {\emph {\bibinfo {title} {Essential Quantum
  Optics: From Quantum Measurements to Black Holes}}}\ (\bibinfo  {publisher}
  {Cambridge University Press},\ \bibinfo {address} {Cambridge, UK},\ \bibinfo
  {year} {2010})\BibitemShut {NoStop}%
\bibitem [{\citenamefont {{Agarwal}}(2013)}]{aga}%
  \BibitemOpen
  \bibfield  {author} {\bibinfo {author} {\bibfnamefont {G.~S.}\ \bibnamefont
  {{Agarwal}}},\ }\href@noop {} {\emph {\bibinfo {title} {{Quantum Optics}}}}\
  (\bibinfo  {publisher} {Cambridge University Press},\ \bibinfo {address}
  {Cambridge, UK},\ \bibinfo {year} {2013})\BibitemShut {NoStop}%
\bibitem [{\citenamefont {Gr\"{o}chenig}(2001)}]{groe}%
  \BibitemOpen
  \bibfield  {author} {\bibinfo {author} {\bibfnamefont {K.}~\bibnamefont
  {Gr\"{o}chenig}},\ }\href@noop {} {\emph {\bibinfo {title} {Foundations of
  Time-Frequency Analysis}}},\ Applied and Numerical Harmonic Analysis\
  (\bibinfo  {publisher} {Springer Science+Business Media},\ \bibinfo {address}
  {New York},\ \bibinfo {year} {2001})\BibitemShut {NoStop}%
\bibitem [{\citenamefont {Cohen}(2013)}]{coh}%
  \BibitemOpen
  \bibfield  {author} {\bibinfo {author} {\bibfnamefont {L.}~\bibnamefont
  {Cohen}},\ }\href@noop {} {\emph {\bibinfo {title} {The Weyl Operator and its
  Generalization}}},\ \bibinfo {series} {Pseudo-Differential Operators, Theory
  and Applications}, Vol.~\bibinfo {volume} {9}\ (\bibinfo  {publisher}
  {Birkh\"{a}user, Springer},\ \bibinfo {address} {Basel},\ \bibinfo {year}
  {2013})\BibitemShut {NoStop}%
\bibitem [{\citenamefont {Kastrup}(2006)}]{ka}%
  \BibitemOpen
  \bibfield  {author} {\bibinfo {author} {\bibfnamefont {H.~A.}\ \bibnamefont
  {Kastrup}},\ }\bibfield  {title} {\enquote {\bibinfo {title} {Quantization of
  the canonically conjugate pair angle and orbital angular momentum},}\
  }\href@noop {} {\bibfield  {journal} {\bibinfo  {journal} {Phys. Rev. A}\
  }\textbf {\bibinfo {volume} {73}},\ \bibinfo {pages} {052104} (\bibinfo
  {year} {2006})},\ \bibinfo {note} {here Ap\-pendix A. Compared to this Ref.
  \cite{ka} the present follow-up paper has some minor changes of
  notations.}\BibitemShut {Stop}%
\bibitem [{\citenamefont {Berry}(1977)}]{ber}%
  \BibitemOpen
  \bibfield  {author} {\bibinfo {author} {\bibfnamefont {M.~V.}\ \bibnamefont
  {Berry}},\ }\bibfield  {title} {\enquote {\bibinfo {title} {Semi-classical
  mechanics in phase space: A study of wigner{\textquoteright}s function},}\
  }\href@noop {} {\bibfield  {journal} {\bibinfo  {journal} {Phil.\ Trans.\
  Roy.\ Soc.\ London A}\ }\textbf {\bibinfo {volume} {287}},\ \bibinfo {pages}
  {237} (\bibinfo {year} {1977})}\BibitemShut {NoStop}%
\bibitem [{\citenamefont {Mukunda}(1979)}]{muk}%
  \BibitemOpen
  \bibfield  {author} {\bibinfo {author} {\bibfnamefont {N.}~\bibnamefont
  {Mukunda}},\ }\bibfield  {title} {\enquote {\bibinfo {title} {Wigner
  distribution for angle coordinates in quantum mechanics},}\ }\href@noop {}
  {\bibfield  {journal} {\bibinfo  {journal} {Am. J. Phys.}\ }\textbf {\bibinfo
  {volume} {47}},\ \bibinfo {pages} {182} (\bibinfo {year} {1979})}\BibitemShut
  {NoStop}%
\bibitem [{\citenamefont {Rigas}\ \emph {et~al.}(2011)\citenamefont {Rigas},
  \citenamefont {Sánchez-Soto}, \citenamefont {Klimov}, \citenamefont
  {Řeháček},\ and\ \citenamefont {Hradil}}]{soto}%
  \BibitemOpen
  \bibfield  {author} {\bibinfo {author} {\bibfnamefont {I.}~\bibnamefont
  {Rigas}}, \bibinfo {author} {\bibfnamefont {L.~L.}\ \bibnamefont
  {Sánchez-Soto}}, \bibinfo {author} {\bibfnamefont {A.~B.}\ \bibnamefont
  {Klimov}}, \bibinfo {author} {\bibfnamefont {J.}~\bibnamefont {Řeháček}},
  \ and\ \bibinfo {author} {\bibfnamefont {Z.}~\bibnamefont {Hradil}},\
  }\bibfield  {title} {\enquote {\bibinfo {title} {Orbital angular momentum in
  phase space},}\ }\href@noop {} {\bibfield  {journal} {\bibinfo  {journal}
  {Ann. Phys. (NY)}\ }\textbf {\bibinfo {volume} {326}},\ \bibinfo {pages}
  {426} (\bibinfo {year} {2011})}\BibitemShut {NoStop}%
\bibitem [{\citenamefont {Przanowski}\ \emph {et~al.}(2014)\citenamefont
  {Przanowski}, \citenamefont {Brzykcy},\ and\ \citenamefont {Tosiek}}]{prza}%
  \BibitemOpen
  \bibfield  {author} {\bibinfo {author} {\bibfnamefont {M.}~\bibnamefont
  {Przanowski}}, \bibinfo {author} {\bibfnamefont {P.}~\bibnamefont {Brzykcy}},
  \ and\ \bibinfo {author} {\bibfnamefont {J.}~\bibnamefont {Tosiek}},\
  }\bibfield  {title} {\enquote {\bibinfo {title} {From the {W}eyl quantization
  of a particle on the circle to number–phase {W}igner functions},}\
  }\href@noop {} {\bibfield  {journal} {\bibinfo  {journal} {Ann. Phys. (NY)}\
  }\textbf {\bibinfo {volume} {351}},\ \bibinfo {pages} {919} (\bibinfo {year}
  {2014})}\BibitemShut {NoStop}%
\bibitem [{\citenamefont {Mukunda}\ \emph {et~al.}(2005)\citenamefont
  {Mukunda}, \citenamefont {Marmo}, \citenamefont {Zampini}, \citenamefont
  {Chaturvedi},\ and\ \citenamefont {Simon}}]{muk2}%
  \BibitemOpen
  \bibfield  {author} {\bibinfo {author} {\bibfnamefont {N.}~\bibnamefont
  {Mukunda}}, \bibinfo {author} {\bibfnamefont {G.}~\bibnamefont {Marmo}},
  \bibinfo {author} {\bibfnamefont {A.}~\bibnamefont {Zampini}}, \bibinfo
  {author} {\bibfnamefont {S.}~\bibnamefont {Chaturvedi}}, \ and\ \bibinfo
  {author} {\bibfnamefont {R.}~\bibnamefont {Simon}},\ }\bibfield  {title}
  {\enquote {\bibinfo {title} {Wigner-{W}eyl isomorphism for quantum mechanics
  on {L}ie groups},}\ }\href@noop {} {\bibfield  {journal} {\bibinfo  {journal}
  {J. Math. Phys.}\ }\textbf {\bibinfo {volume} {46}},\ \bibinfo {pages}
  {012106} (\bibinfo {year} {2005})}\BibitemShut {NoStop}%
\bibitem [{\citenamefont {Fronsdal}(1979)}]{fron}%
  \BibitemOpen
  \bibfield  {author} {\bibinfo {author} {\bibfnamefont {C.}~\bibnamefont
  {Fronsdal}},\ }\bibfield  {title} {\enquote {\bibinfo {title} {Some ideas
  about quantization},}\ }\href@noop {} {\bibfield  {journal} {\bibinfo
  {journal} {Rep. Math. Phys.}\ }\textbf {\bibinfo {volume} {15}},\ \bibinfo
  {pages} {111} (\bibinfo {year} {1979})}\BibitemShut {NoStop}%
\bibitem [{\citenamefont {Gadella}\ \emph {et~al.}(1991)\citenamefont
  {Gadella}, \citenamefont {Martin}, \citenamefont {Nieto},\ and\ \citenamefont
  {del Olmo}}]{gad}%
  \BibitemOpen
  \bibfield  {author} {\bibinfo {author} {\bibfnamefont {M.}~\bibnamefont
  {Gadella}}, \bibinfo {author} {\bibfnamefont {M.~A.}\ \bibnamefont {Martin}},
  \bibinfo {author} {\bibfnamefont {L.~M.}\ \bibnamefont {Nieto}}, \ and\
  \bibinfo {author} {\bibfnamefont {M.~A.}\ \bibnamefont {del Olmo}},\
  }\bibfield  {title} {\enquote {\bibinfo {title} {The {S}tratonovich–{W}eyl
  correspondence for one-dimensional kinematical groups},}\ }\href@noop {}
  {\bibfield  {journal} {\bibinfo  {journal} {J. Math. Phys.}\ }\textbf
  {\bibinfo {volume} {32}},\ \bibinfo {pages} {1182} (\bibinfo {year}
  {1991})}\BibitemShut {NoStop}%
\bibitem [{\citenamefont {Arratia}\ and\ \citenamefont {del
  Olmo}(1997)}]{arra}%
  \BibitemOpen
  \bibfield  {author} {\bibinfo {author} {\bibfnamefont {O.}~\bibnamefont
  {Arratia}}\ and\ \bibinfo {author} {\bibfnamefont {M.~A.}\ \bibnamefont {del
  Olmo}},\ }\bibfield  {title} {\enquote {\bibinfo {title} {Moyal quantization
  on the cylinder},}\ }\href@noop {} {\bibfield  {journal} {\bibinfo  {journal}
  {Rep. Math. Phys.}\ }\textbf {\bibinfo {volume} {40}},\ \bibinfo {pages}
  {149} (\bibinfo {year} {1997})}\BibitemShut {NoStop}%
\bibitem [{\citenamefont {Isham}(1984)}]{ish}%
  \BibitemOpen
  \bibfield  {author} {\bibinfo {author} {\bibfnamefont {C.~J.}\ \bibnamefont
  {Isham}},\ }\enquote {\bibinfo {title} {Topological and global aspects of
  quantum theory},}\ in\ \href@noop {} {\emph {\bibinfo {booktitle}
  {Relativity, Groups and Topology II}}},\ \bibinfo {series and number} {Les
  Houches Session XL, 1983},\ \bibinfo {editor} {edited by\ \bibinfo {editor}
  {\bibfnamefont {B.~S.}\ \bibnamefont {DeWitt}}\ and\ \bibinfo {editor}
  {\bibfnamefont {R.}~\bibnamefont {Stora}}}\ (\bibinfo  {publisher} {North
  Holland},\ \bibinfo {address} {Amsterdam},\ \bibinfo {year} {1984})\ pp.\
  \bibinfo {pages} {1059 -- 1290},\ \bibinfo {note} {here especially pp.\
  1170--1176, 1224--1226}\BibitemShut {NoStop}%
\bibitem [{\citenamefont {Wolf}(1996)}]{wol1}%
  \BibitemOpen
  \bibfield  {author} {\bibinfo {author} {\bibfnamefont {K.~B.}\ \bibnamefont
  {Wolf}},\ }\bibfield  {title} {\enquote {\bibinfo {title} {Wigner
  distribution function for paraxial polychromatic optics},}\ }\href@noop {}
  {\bibfield  {journal} {\bibinfo  {journal} {Opt. Comm.}\ }\textbf {\bibinfo
  {volume} {132}},\ \bibinfo {pages} {343} (\bibinfo {year}
  {1996})}\BibitemShut {NoStop}%
\bibitem [{\citenamefont {Nieto}\ \emph {et~al.}(1998)\citenamefont {Nieto},
  \citenamefont {Atakishiyev}, \citenamefont {Chumakov},\ and\ \citenamefont
  {Wolf}}]{wol2}%
  \BibitemOpen
  \bibfield  {author} {\bibinfo {author} {\bibfnamefont {L.~M.}\ \bibnamefont
  {Nieto}}, \bibinfo {author} {\bibfnamefont {N.~M.}\ \bibnamefont
  {Atakishiyev}}, \bibinfo {author} {\bibfnamefont {S.~M.}\ \bibnamefont
  {Chumakov}}, \ and\ \bibinfo {author} {\bibfnamefont {K.~B.}\ \bibnamefont
  {Wolf}},\ }\bibfield  {title} {\enquote {\bibinfo {title} {Wigner
  distribution function for {E}uclidean systems},}\ }\href@noop {} {\bibfield
  {journal} {\bibinfo  {journal} {J. Phys. A: Math. Gen.}\ }\textbf {\bibinfo
  {volume} {31}},\ \bibinfo {pages} {3875} (\bibinfo {year}
  {1998})}\BibitemShut {NoStop}%
\bibitem [{\citenamefont {Ali}\ \emph {et~al.}(2000)\citenamefont {Ali},
  \citenamefont {Atakishiyev}, \citenamefont {Chumakov},\ and\ \citenamefont
  {Wolf}}]{wol3}%
  \BibitemOpen
  \bibfield  {author} {\bibinfo {author} {\bibfnamefont {S.~T.}\ \bibnamefont
  {Ali}}, \bibinfo {author} {\bibfnamefont {N.~M.}\ \bibnamefont
  {Atakishiyev}}, \bibinfo {author} {\bibfnamefont {S.~M.}\ \bibnamefont
  {Chumakov}}, \ and\ \bibinfo {author} {\bibfnamefont {K.~B.}\ \bibnamefont
  {Wolf}},\ }\bibfield  {title} {\enquote {\bibinfo {title} {The {W}igner
  function for general {L}ie groups and the wavelet transform},}\ }\href@noop
  {} {\bibfield  {journal} {\bibinfo  {journal} {Ann. Henri Poincar\'e}\
  }\textbf {\bibinfo {volume} {1}},\ \bibinfo {pages} {685} (\bibinfo {year}
  {2000})}\BibitemShut {NoStop}%
\bibitem [{\citenamefont {Moyal}(1949)}]{moy}%
  \BibitemOpen
  \bibfield  {author} {\bibinfo {author} {\bibfnamefont {J.~E.}\ \bibnamefont
  {Moyal}},\ }\bibfield  {title} {\enquote {\bibinfo {title} {Quantum mechanics
  as a statistical theory},}\ }\href@noop {} {\bibfield  {journal} {\bibinfo
  {journal} {Proc. Cambridge Philos. Soc.}\ }\textbf {\bibinfo {volume} {45}},\
  \bibinfo {pages} {99} (\bibinfo {year} {1949})}\BibitemShut {NoStop}%
\bibitem [{\citenamefont {Folland}(1989)}]{fol}%
  \BibitemOpen
  \bibfield  {author} {\bibinfo {author} {\bibfnamefont {G.~B.}\ \bibnamefont
  {Folland}},\ }\href@noop {} {\emph {\bibinfo {title} {Harmonic Analysis in
  Phase Space}}},\ \bibinfo {series} {The Annals of Mathematics Studies}, Vol.\
  \bibinfo {volume} {122}\ (\bibinfo  {publisher} {Princeton University
  Press},\ \bibinfo {address} {Princeton, NJ},\ \bibinfo {year} {1989})\
  \bibinfo {note} {{C}hs.\ 1 and 2}\BibitemShut {NoStop}%
\bibitem [{\citenamefont {de~Gosson}(2006)}]{gos}%
  \BibitemOpen
  \bibfield  {author} {\bibinfo {author} {\bibfnamefont {M.}~\bibnamefont
  {de~Gosson}},\ }\href@noop {} {\emph {\bibinfo {title} {Symplectic Geometry
  and Quantum Mechanics}}},\ \bibinfo {series} {Operator Theory: Advances and
  Applications}, Vol.\ \bibinfo {volume} {166}\ (\bibinfo  {publisher}
  {Birkh\"{a}user},\ \bibinfo {address} {Basel},\ \bibinfo {year} {2006})\
  \bibinfo {note} {{P}arts II and III}\BibitemShut {NoStop}%
\bibitem [{\citenamefont {Whittaker}(1915)}]{whi}%
  \BibitemOpen
  \bibfield  {author} {\bibinfo {author} {\bibfnamefont {E.~T.}\ \bibnamefont
  {Whittaker}},\ }\bibfield  {title} {\enquote {\bibinfo {title} {On the
  functions which are represented by the expansions of the
  interpolation-theory},}\ }\href@noop {} {\bibfield  {journal} {\bibinfo
  {journal} {Proc. Roy. Soc. Edinburgh}\ }\textbf {\bibinfo {volume} {35}},\
  \bibinfo {pages} {181} (\bibinfo {year} {1915})}\BibitemShut {NoStop}%
\bibitem [{\citenamefont {McNamee}\ \emph {et~al.}(1971)\citenamefont
  {McNamee}, \citenamefont {Stenger},\ and\ \citenamefont {Whitney}}]{mcn}%
  \BibitemOpen
  \bibfield  {author} {\bibinfo {author} {\bibfnamefont {J.}~\bibnamefont
  {McNamee}}, \bibinfo {author} {\bibfnamefont {F.}~\bibnamefont {Stenger}}, \
  and\ \bibinfo {author} {\bibfnamefont {E.~L.}\ \bibnamefont {Whitney}},\
  }\bibfield  {title} {\enquote {\bibinfo {title} {Whittak\-er's cardinal
  function in retrospect},}\ }\href@noop {} {\bibfield  {journal} {\bibinfo
  {journal} {Mathema\-tics of Computation}\ }\textbf {\bibinfo {volume} {25}},\
  \bibinfo {pages} {141} (\bibinfo {year} {1971})}\BibitemShut {NoStop}%
\bibitem [{\citenamefont {Stenger}(1981)}]{ste1}%
  \BibitemOpen
  \bibfield  {author} {\bibinfo {author} {\bibfnamefont {F.}~\bibnamefont
  {Stenger}},\ }\bibfield  {title} {\enquote {\bibinfo {title} {Numerical
  methods based on {W}hittaker cardinal, or sinc functions},}\ }\href@noop {}
  {\bibfield  {journal} {\bibinfo  {journal} {SIAM Rev.}\ }\textbf {\bibinfo
  {volume} {23}},\ \bibinfo {pages} {165} (\bibinfo {year} {1981})}\BibitemShut
  {NoStop}%
\bibitem [{\citenamefont {Butzer}(1983)}]{bu}%
  \BibitemOpen
  \bibfield  {author} {\bibinfo {author} {\bibfnamefont {P.~L.}\ \bibnamefont
  {Butzer}},\ }\bibfield  {title} {\enquote {\bibinfo {title} {A survey of the
  {W}hittaker-{S}hannon sampling theorem and some of its extensions},}\
  }\href@noop {} {\bibfield  {journal} {\bibinfo  {journal} {J. Mathem.
  Research and Exposition (now: ... and Application)}\ }\textbf {\bibinfo
  {volume} {3}},\ \bibinfo {pages} {185} (\bibinfo {year} {1983})},\ \bibinfo
  {note} {publ. by Dalian Univ. of Technology and China Soc. for Industrial and
  Appl. Mathem.; journal available on the web}\BibitemShut {NoStop}%
\bibitem [{\citenamefont {Higgins}(1985)}]{hig}%
  \BibitemOpen
  \bibfield  {author} {\bibinfo {author} {\bibfnamefont {J.~R.}\ \bibnamefont
  {Higgins}},\ }\bibfield  {title} {\enquote {\bibinfo {title} {Five short
  stories about the cardinal series},}\ }\href@noop {} {\bibfield  {journal}
  {\bibinfo  {journal} {Bull. (N.S.) Amer. Math. Soc.}\ }\textbf {\bibinfo
  {volume} {12}},\ \bibinfo {pages} {45} (\bibinfo {year} {1985})}\BibitemShut
  {NoStop}%
\bibitem [{\citenamefont {Stenger}(1993)}]{ste2}%
  \BibitemOpen
  \bibfield  {author} {\bibinfo {author} {\bibfnamefont {Frank}\ \bibnamefont
  {Stenger}},\ }\href@noop {} {\emph {\bibinfo {title} {Numerical Methods Based
  on Sinc and Analytic Functions}}},\ \bibinfo {series} {Springer Series in
  Computational Mathematics}, Vol.~\bibinfo {volume} {20}\ (\bibinfo
  {publisher} {Springer-Verlag},\ \bibinfo {address} {New York etc.},\ \bibinfo
  {year} {1993})\BibitemShut {NoStop}%
\bibitem [{\citenamefont {Stenger}(2011)}]{ste}%
  \BibitemOpen
  \bibfield  {author} {\bibinfo {author} {\bibfnamefont {F.}~\bibnamefont
  {Stenger}},\ }\href@noop {} {\emph {\bibinfo {title} {Handbook of Sinc
  Numerical Methods}}},\ Chapman \& Hall/CRC Numerical Analysis and Scientific
  Computing\ (\bibinfo  {publisher} {CRC Press Taylor \& Francis Group},\
  \bibinfo {address} {Boca Raton, FL},\ \bibinfo {year} {2011})\BibitemShut
  {NoStop}%
\bibitem [{\citenamefont {Vetterli}\ \emph {et~al.}(2014)\citenamefont
  {Vetterli}, \citenamefont {Kova\v{c}evi\'{c}},\ and\ \citenamefont
  {Goyal}}]{vet}%
  \BibitemOpen
  \bibfield  {author} {\bibinfo {author} {\bibfnamefont {M.}~\bibnamefont
  {Vetterli}}, \bibinfo {author} {\bibfnamefont {J.}~\bibnamefont
  {Kova\v{c}evi\'{c}}}, \ and\ \bibinfo {author} {\bibfnamefont {V.~K.}\
  \bibnamefont {Goyal}},\ }\href@noop {} {\emph {\bibinfo {title} {Foundations
  of Signal Processing}}}\ (\bibinfo  {publisher} {Cambridge University
  Press},\ \bibinfo {address} {Cambridge, UK},\ \bibinfo {year}
  {2014})\BibitemShut {NoStop}%
\bibitem [{\citenamefont {Eldar}(2015)}]{eld}%
  \BibitemOpen
  \bibfield  {author} {\bibinfo {author} {\bibfnamefont {Y.~C.}\ \bibnamefont
  {Eldar}},\ }\href@noop {} {\emph {\bibinfo {title} {Sampling Theory}}}\
  (\bibinfo  {publisher} {Cambridge University Press},\ \bibinfo {address}
  {Cambridge, UK},\ \bibinfo {year} {2015})\BibitemShut {NoStop}%
\bibitem [{\citenamefont {Louisell}(1963)}]{louis}%
  \BibitemOpen
  \bibfield  {author} {\bibinfo {author} {\bibfnamefont {W.~H.}\ \bibnamefont
  {Louisell}},\ }\bibfield  {title} {\enquote {\bibinfo {title} {Amplitude and
  phase uncertainty relations},}\ }\href@noop {} {\bibfield  {journal}
  {\bibinfo  {journal} {Phys. Lett.}\ }\textbf {\bibinfo {volume} {7}},\
  \bibinfo {pages} {60} (\bibinfo {year} {1963})}\BibitemShut {NoStop}%
\bibitem [{\citenamefont {Mackey}(1963)}]{mack}%
  \BibitemOpen
  \bibfield  {author} {\bibinfo {author} {\bibfnamefont {G.~W.}\ \bibnamefont
  {Mackey}},\ }\href@noop {} {\emph {\bibinfo {title} {{The Mathematical
  Foundations of Quantum Mechanics}}}}\ (\bibinfo  {publisher} {W.A. Benjamin,
  Inc.},\ \bibinfo {address} {New York},\ \bibinfo {year} {1963})\ \bibinfo
  {note} {p.\ 103}\BibitemShut {NoStop}%
\bibitem [{\citenamefont {Vilenkin}(1968)}]{vil}%
  \BibitemOpen
  \bibfield  {author} {\bibinfo {author} {\bibfnamefont {N.~I.}\ \bibnamefont
  {Vilenkin}},\ }\href@noop {} {\emph {\bibinfo {title} {{Special Functions and
  the Theory of Group Representations}}}},\ \bibinfo {series} {Translations of
  Mathematical Monographs}, Vol.~\bibinfo {volume} {22}\ (\bibinfo  {publisher}
  {Amer. Math. Soc.},\ \bibinfo {address} {Providence, RI},\ \bibinfo {year}
  {1968})\ \bibinfo {note} {{C}h.\ IV}\BibitemShut {NoStop}%
\bibitem [{\citenamefont {Sugiura}(1990)}]{sug}%
  \BibitemOpen
  \bibfield  {author} {\bibinfo {author} {\bibfnamefont {M.}~\bibnamefont
  {Sugiura}},\ }\href@noop {} {\emph {\bibinfo {title} {Unitary Representations
  and Harmonic Ana\-lysis: An Introduction}}},\ \bibinfo {series}
  {North-Holland Mathematical Library}, Vol.~\bibinfo {volume} {44}\ (\bibinfo
  {publisher} {Elsevier},\ \bibinfo {address} {Amsterdam},\ \bibinfo {year}
  {1990})\ \bibinfo {note} {{C}h.\ IV}\BibitemShut {NoStop}%
\bibitem [{\citenamefont {Chirikjian}\ and\ \citenamefont
  {Kyatkin}(2000)}]{chi}%
  \BibitemOpen
  \bibfield  {author} {\bibinfo {author} {\bibfnamefont {G.~S.}\ \bibnamefont
  {Chirikjian}}\ and\ \bibinfo {author} {\bibfnamefont {A.~B.}\ \bibnamefont
  {Kyatkin}},\ }\href@noop {} {\emph {\bibinfo {title} {Engineering
  Applications of Noncommutative Harmonic Analysis, with Emphasis on Rotation
  and Motion Groups}}}\ (\bibinfo  {publisher} {CRC Press},\ \bibinfo {address}
  {Boca Raton, FL},\ \bibinfo {year} {2000})\ \bibinfo {note} {p.\ 151, (of
  general interest for the present paper is especially Ch.\ 10)}\BibitemShut
  {NoStop}%
\bibitem [{\citenamefont {Whittaker}\ and\ \citenamefont
  {Watson}(1969)}]{bess}%
  \BibitemOpen
  \bibfield  {author} {\bibinfo {author} {\bibfnamefont {E.~T.}\ \bibnamefont
  {Whittaker}}\ and\ \bibinfo {author} {\bibfnamefont {G.~N.}\ \bibnamefont
  {Watson}},\ }\href@noop {} {\emph {\bibinfo {title} {A Course of Modern
  Analysis}}},\ \bibinfo {edition} {4th}\ ed.\ (\bibinfo  {publisher}
  {Cambridge University Press},\ \bibinfo {address} {Cambridge, UK},\ \bibinfo
  {year} {1969})\ p.\ \bibinfo {pages} {362}\BibitemShut {NoStop}%
\bibitem [{\citenamefont {Morse}\ and\ \citenamefont {Feshbach}(1953)}]{mor}%
  \BibitemOpen
  \bibfield  {author} {\bibinfo {author} {\bibfnamefont {Ph.~M.}\ \bibnamefont
  {Morse}}\ and\ \bibinfo {author} {\bibfnamefont {H.}~\bibnamefont
  {Feshbach}},\ }\href@noop {} {\emph {\bibinfo {title} {Methods of Theoretical
  Physics, Vol. I}}},\ International Series in Pure and Applied Physics\
  (\bibinfo  {publisher} {McGraw-Hill Book Co., Inc.},\ \bibinfo {address} {New
  York},\ \bibinfo {year} {1953})\ p.\ \bibinfo {pages} {766},\ \bibinfo {note}
  {formula (6.3.62)}\BibitemShut {NoStop}%
\bibitem [{\citenamefont {Prudnikov}\ \emph {et~al.}(1986)\citenamefont
  {Prudnikov}, \citenamefont {Brychkov},\ and\ \citenamefont {Marichev}}]{pru}%
  \BibitemOpen
  \bibfield  {author} {\bibinfo {author} {\bibfnamefont {A.~P.}\ \bibnamefont
  {Prudnikov}}, \bibinfo {author} {\bibfnamefont {Yu.~A.}\ \bibnamefont
  {Brychkov}}, \ and\ \bibinfo {author} {\bibfnamefont {O.~I.}\ \bibnamefont
  {Marichev}},\ }\href@noop {} {\emph {\bibinfo {title} {Integrals and
  Series}}},\ Vol.~\bibinfo {volume} {1}\ (\bibinfo  {publisher} {Gordon and
  Breach Science Publishers},\ \bibinfo {address} {New York, London etc.},\
  \bibinfo {year} {1986})\ \bibinfo {note} {p. 727, formula 13.6.}\BibitemShut
  {Stop}%
\bibitem [{\citenamefont {de~Groot}\ and\ \citenamefont
  {Suttorp}(1972)}]{groo}%
  \BibitemOpen
  \bibfield  {author} {\bibinfo {author} {\bibfnamefont {S.~R.}\ \bibnamefont
  {de~Groot}}\ and\ \bibinfo {author} {\bibfnamefont {L.~G.}\ \bibnamefont
  {Suttorp}},\ }\href@noop {} {\emph {\bibinfo {title} {Foundations of
  Electrodynamics}}}\ (\bibinfo  {publisher} {North-Holland},\ \bibinfo
  {address} {Amsterdam},\ \bibinfo {year} {1972})\ \bibinfo {note} {{C}h.\ VI
  (the representation here relies on Refs.\ \cite{leaf1, leaf2}).}\BibitemShut
  {Stop}%
\bibitem [{\citenamefont {Hardy}(1941)}]{har}%
  \BibitemOpen
  \bibfield  {author} {\bibinfo {author} {\bibfnamefont {G.~H.}\ \bibnamefont
  {Hardy}},\ }\bibfield  {title} {\enquote {\bibinfo {title} {Notes on special
  systems of orthogonal functions ({IV}): The orthogonal functions of
  {W}hittaker's series},}\ }\href@noop {} {\bibfield  {journal} {\bibinfo
  {journal} {Proc. Cambridge Philos. Soc.}\ }\textbf {\bibinfo {volume} {37}},\
  \bibinfo {pages} {331} (\bibinfo {year} {1941})},\ \bibinfo {note} {reprinted
  in {\it Collected Papers of G.H. Hardy, vol. III} (Clarendon Press, Oxford,
  1969) p.\ 466}\BibitemShut {NoStop}%
\bibitem [{\citenamefont {Christensen}(2008)}]{chr}%
  \BibitemOpen
  \bibfield  {author} {\bibinfo {author} {\bibfnamefont {O.}~\bibnamefont
  {Christensen}},\ }\href@noop {} {\emph {\bibinfo {title} {Frames and Bases,
  An Introductory Course}}},\ Applied and Numerical Harmonic Analysis\
  (\bibinfo  {publisher} {Birkh\"{a}user},\ \bibinfo {address} {Boston},\
  \bibinfo {year} {2008})\ \bibinfo {note} {here Ch. 3.8}\BibitemShut {NoStop}%
\bibitem [{sch()}]{schl1}%
  \BibitemOpen
  \href@noop {} {}\bibinfo {note} {See, e.g.\ Ref.\ \cite{schl}, Ch. 4 or Ref.\
  \cite{aga}, Chs. 1.7 and 1.8.}\BibitemShut {Stop}%
\bibitem [{\citenamefont {\v{R}ehá\v{c}ek}\ \emph {et~al.}(2008)\citenamefont
  {\v{R}ehá\v{c}ek}, \citenamefont {Bouchal}, \citenamefont
  {\v{C}elechovský}, \citenamefont {Hradil},\ and\ \citenamefont
  {Sánchez-Soto}}]{soto1}%
  \BibitemOpen
  \bibfield  {author} {\bibinfo {author} {\bibfnamefont {J.}~\bibnamefont
  {\v{R}ehá\v{c}ek}}, \bibinfo {author} {\bibfnamefont {Z.}~\bibnamefont
  {Bouchal}}, \bibinfo {author} {\bibfnamefont {R.}~\bibnamefont
  {\v{C}elechovský}}, \bibinfo {author} {\bibfnamefont {Z.}~\bibnamefont
  {Hradil}}, \ and\ \bibinfo {author} {\bibfnamefont {L.~L.}\ \bibnamefont
  {Sánchez-Soto}},\ }\bibfield  {title} {\enquote {\bibinfo {title}
  {Experimental test of uncertainty relations for quantum mechanics on a
  circle},}\ }\href@noop {} {\bibfield  {journal} {\bibinfo  {journal} {Phys.
  Rev. A}\ ,\ \bibinfo {pages} {032110}} (\bibinfo {year} {2008})}\BibitemShut
  {NoStop}%
\bibitem [{\citenamefont {Robertson}(1929)}]{rob}%
  \BibitemOpen
  \bibfield  {author} {\bibinfo {author} {\bibfnamefont {H.~P.}\ \bibnamefont
  {Robertson}},\ }\bibfield  {title} {\enquote {\bibinfo {title} {The
  uncertainty principle},}\ }\href@noop {} {\bibfield  {journal} {\bibinfo
  {journal} {Phys. Rev.}\ }\textbf {\bibinfo {volume} {34}},\ \bibinfo {pages}
  {163} (\bibinfo {year} {1929})}\BibitemShut {NoStop}%
\bibitem [{\citenamefont {Schr\"odinger}(1930)}]{schr}%
  \BibitemOpen
  \bibfield  {author} {\bibinfo {author} {\bibfnamefont {E.}~\bibnamefont
  {Schr\"odinger}},\ }\bibfield  {title} {\enquote {\bibinfo {title} {Zum
  {H}eisenbergschen {U}nsch\"{a}rfeprin\-zip},}\ }\href@noop {} {\bibfield
  {journal} {\bibinfo  {journal} {Sitzungsber. Preuss. Akad. Wiss., Phys.-math.
  Klas\-se}\ }\textbf {\bibinfo {volume} {19}},\ \bibinfo {pages} {296}
  (\bibinfo {year} {1930})},\ \bibinfo {note} {reprinted in E. Schr\"odinger,
  {\em Collected Papers, Vol. 3}, p.\ 348 (\"{O}sterr. Akad. Wiss. and Friedr.
  Vieweg \& Sohn, Braunschweig/Wiesbaden, Vienna, 1984)}\BibitemShut {NoStop}%
\bibitem [{\citenamefont {Jackiw}(1968)}]{jack}%
  \BibitemOpen
  \bibfield  {author} {\bibinfo {author} {\bibfnamefont {R.}~\bibnamefont
  {Jackiw}},\ }\bibfield  {title} {\enquote {\bibinfo {title} {Minimum
  uncertainty product, number-phase uncertainty product, and coherent
  states},}\ }\href@noop {} {\bibfield  {journal} {\bibinfo  {journal} {J.
  Math. Phys.}\ }\textbf {\bibinfo {volume} {9}},\ \bibinfo {pages} {339}
  (\bibinfo {year} {1968})}\BibitemShut {NoStop}%
\bibitem [{\citenamefont {Forbes}\ \emph {et~al.}(2011)\citenamefont {Forbes},
  \citenamefont {Evans},\ and\ \citenamefont {Hastings}}]{for}%
  \BibitemOpen
  \bibfield  {author} {\bibinfo {author} {\bibfnamefont {C.}~\bibnamefont
  {Forbes}}, \bibinfo {author} {\bibfnamefont {M.}~\bibnamefont {Evans}}, \
  and\ \bibinfo {author} {\bibfnamefont {N.}~\bibnamefont {Hastings}},\
  }\href@noop {} {\emph {\bibinfo {title} {Statistical Distri\-butions}}},\
  \bibinfo {edition} {4th}\ ed.\ (\bibinfo  {publisher} {John Wiley \& Sons,
  Inc.},\ \bibinfo {address} {Hoboken, NJ, USA},\ \bibinfo {year} {2011})\
  \bibinfo {note} {ch.\ 45}\BibitemShut {NoStop}%
\bibitem [{\citenamefont {Mardia}\ and\ \citenamefont {Jupp}(2000)}]{mar}%
  \BibitemOpen
  \bibfield  {author} {\bibinfo {author} {\bibfnamefont {K.~V.}\ \bibnamefont
  {Mardia}}\ and\ \bibinfo {author} {\bibfnamefont {P.~E.}\ \bibnamefont
  {Jupp}},\ }\href@noop {} {\emph {\bibinfo {title} {Directional
  Statistics}}},\ Wiley Series in Probability and Statistics\ (\bibinfo
  {publisher} {John Wiley \& Sons, Ltd},\ \bibinfo {address} {Chichester etc.,
  UK},\ \bibinfo {year} {2000})\ \bibinfo {note} {especially Subsec.\
  3.5.4}\BibitemShut {NoStop}%
\bibitem [{\citenamefont {Watson}(1966)}]{wat}%
  \BibitemOpen
  \bibfield  {author} {\bibinfo {author} {\bibfnamefont {G.~N.}\ \bibnamefont
  {Watson}},\ }\href@noop {} {\emph {\bibinfo {title} {A Treatise on the Theory
  of Bessel Functions}}},\ \bibinfo {edition} {2nd}\ ed.\ (\bibinfo
  {publisher} {Cambridge University Press},\ \bibinfo {address} {Cambridge,
  UK},\ \bibinfo {year} {1966})\ p.\ \bibinfo {pages} {181},\ \bibinfo {note}
  {formula (4)}\BibitemShut {NoStop}%
\bibitem [{wa2()}]{wa2}%
  \BibitemOpen
  \href@noop {} {}\bibinfo {note} {See, e.g.\ Ref.\ \cite{wat}, p.\
  698}\BibitemShut {NoStop}%
\bibitem [{\citenamefont {Erdélyi}(1954)}]{erd}%
  \BibitemOpen
  \bibinfo {editor} {\bibfnamefont {A.}~\bibnamefont {Erdélyi}},\ ed.,\
  \href@noop {} {\emph {\bibinfo {title} {Tables of Integral Transforms}}},\
  Vol.~\bibinfo {volume} {I}\ (\bibinfo  {publisher} {McGraw-Hill Book Co.,
  Inc.},\ \bibinfo {address} {New York etc.},\ \bibinfo {year} {1954})\
  p.~\bibinfo {pages} {59},\ \bibinfo {note} {formula (61)}\BibitemShut
  {NoStop}%
\bibitem [{\citenamefont {Gradshteyn}\ and\ \citenamefont
  {Ryzhik}(1965)}]{gra}%
  \BibitemOpen
  \bibfield  {author} {\bibinfo {author} {\bibfnamefont {I.~S.}\ \bibnamefont
  {Gradshteyn}}\ and\ \bibinfo {author} {\bibfnamefont {I.~M.}\ \bibnamefont
  {Ryzhik}},\ }\href@noop {} {\emph {\bibinfo {title} {Table of Integrals,
  Series and Products}}},\ \bibinfo {edition} {4th}\ ed.\ (\bibinfo
  {publisher} {Academic Press},\ \bibinfo {address} {New York},\ \bibinfo
  {year} {1965})\ \bibinfo {note} {p. 738, formula 6.681, 3.; the number of the
  formula is the same in later editions.}\BibitemShut {Stop}%
\bibitem [{the()}]{thet}%
  \BibitemOpen
  \href@noop {} {}\bibinfo {note} {For the literature on the $\vt$ functions
  see Appendix C of Ref.\ \cite{ka}.}\BibitemShut {Stop}%
\bibitem [{pru()}]{pru2}%
  \BibitemOpen
  \href@noop {} {}\bibinfo {note} {Ref.\ \cite{pru}, p.\ 403, formula
  41}\BibitemShut {NoStop}%
\bibitem [{gra()}]{gra2}%
  \BibitemOpen
  \href@noop {} {}\bibinfo {note} {Ref.\ \cite{gra}, p.\ 480, formula 3.896,
  4.}\BibitemShut {Stop}%
\bibitem [{chi()}]{chi1}%
  \BibitemOpen
  \href@noop {} {}\bibinfo {note} {See, e.g.\ Ref.\ \cite{chi}, Chs.\ 5 and
  6}\BibitemShut {NoStop}%
\bibitem [{\citenamefont {Marsden}\ and\ \citenamefont {Ratiu}(2003)}]{mars}%
  \BibitemOpen
  \bibfield  {author} {\bibinfo {author} {\bibfnamefont {J.~E.}\ \bibnamefont
  {Marsden}}\ and\ \bibinfo {author} {\bibfnamefont {T.~S.}\ \bibnamefont
  {Ratiu}},\ }\href@noop {} {\emph {\bibinfo {title} {Introduction to Mechanics
  and Symmetry}}},\ \bibinfo {edition} {2nd}\ ed.\ (\bibinfo  {publisher}
  {Springer-Verlag},\ \bibinfo {address} {New York},\ \bibinfo {year} {2003})\
  \bibinfo {note} {here Ch.\ 15}\BibitemShut {NoStop}%
\bibitem [{ish()}]{ish1}%
  \BibitemOpen
  \href@noop {} {}\bibinfo {note} {Ref.\ \cite{ish}, example 4.9 (p. 1194); For
  irreducible unitary representations of Euclidean groups $E(n)$ see, e.g.\ Ch.
  XI of Ref.\ \cite{vil} and Ch. 10 of Ref.\ \cite{chi}.}\BibitemShut {Stop}%
\bibitem [{\citenamefont {Leaf}(1968{\natexlab{a}})}]{leaf1}%
  \BibitemOpen
  \bibfield  {author} {\bibinfo {author} {\bibfnamefont {B.}~\bibnamefont
  {Leaf}},\ }\bibfield  {title} {\enquote {\bibinfo {title} {Weyl
  transformation and the classical limit of quantum mechanics},}\ }\href@noop
  {} {\bibfield  {journal} {\bibinfo  {journal} {J. Math. Phys.}\ }\textbf
  {\bibinfo {volume} {9}},\ \bibinfo {pages} {65} (\bibinfo {year}
  {1968}{\natexlab{a}})}\BibitemShut {NoStop}%
\bibitem [{\citenamefont {Leaf}(1968{\natexlab{b}})}]{leaf2}%
  \BibitemOpen
  \bibfield  {author} {\bibinfo {author} {\bibfnamefont {B.}~\bibnamefont
  {Leaf}},\ }\bibfield  {title} {\enquote {\bibinfo {title} {Weyl transform in
  nonrelativistic quantum dynamics},}\ }\href@noop {} {\bibfield  {journal}
  {\bibinfo  {journal} {J. Math. Phys.}\ }\textbf {\bibinfo {volume} {9}},\
  \bibinfo {pages} {769} (\bibinfo {year} {1968}{\natexlab{b}})}\BibitemShut
  {NoStop}%
\bibitem [{\citenamefont {Alonso}\ \emph {et~al.}(2003)\citenamefont {Alonso},
  \citenamefont {Pogosyan},\ and\ \citenamefont {Wolf}}]{wol4}%
  \BibitemOpen
  \bibfield  {author} {\bibinfo {author} {\bibfnamefont {M.~A.}\ \bibnamefont
  {Alonso}}, \bibinfo {author} {\bibfnamefont {G.~S.}\ \bibnamefont
  {Pogosyan}}, \ and\ \bibinfo {author} {\bibfnamefont {K.~B.}\ \bibnamefont
  {Wolf}},\ }\bibfield  {title} {\enquote {\bibinfo {title} {Wigner functions
  for curved spaces. ii. on spheres},}\ }\href@noop {} {\bibfield  {journal}
  {\bibinfo  {journal} {J. Math. Phys.}\ }\textbf {\bibinfo {volume} {44}},\
  \bibinfo {pages} {1472} (\bibinfo {year} {2003})},\ \bibinfo {note} {here
  Subsec. IV. A.}\BibitemShut {Stop}%
\end{thebibliography}%
\end{document}